\documentclass[aps,prb,superscriptaddress,twocolumn,longbibliography]{revtex4-2}

\usepackage{graphicx}
\usepackage{dcolumn}
\usepackage{bm}
\usepackage{siunitx}

\begin{document}

\preprint{APS/123-QED}

\title{Detection of hidden gratings through multilayer nanostructures using light and sound }

\author{Stephen Edward}
\email{sedward@arcnl.nl}
\affiliation{Advanced Research Center for Nanolithography, Science Park 106, 1098 XG Amsterdam, the Netherlands}
\affiliation{Universiteit van Amsterdam, Science Park 904, 1098 XH Amsterdam, the Netherlands}

\author{Hao Zhang}

\affiliation{Advanced Research Center for Nanolithography, Science Park 106, 1098 XG Amsterdam, the Netherlands}
\affiliation{Vrije Universiteit Amsterdam, De Boelelaan 1105, 1081 HV, Amsterdam, the Netherlands}

\author{Irwan Setija}
\affiliation{ASML Research, De Run 6501, 5504 DR Veldhoven, the Netherlands}

\author{Vanessa Verrina}
\affiliation{Advanced Research Center for Nanolithography, Science Park 106, 1098 XG Amsterdam, the Netherlands}
\affiliation{Universiteit van Amsterdam, Science Park 904, 1098 XH Amsterdam, the Netherlands}

\author{Alessandro Antoncecchi}
\affiliation{Advanced Research Center for Nanolithography, Science Park 106, 1098 XG Amsterdam, the Netherlands}
\affiliation{Vrije Universiteit Amsterdam, De Boelelaan 1105, 1081 HV, Amsterdam, the Netherlands}

\author{Stefan Witte}
\affiliation{Advanced Research Center for Nanolithography, Science Park 106, 1098 XG Amsterdam, the Netherlands}
\affiliation{Vrije Universiteit Amsterdam, De Boelelaan 1105, 1081 HV, Amsterdam, the Netherlands}

\author{Paul Planken}
\email{planken@arcnl.nl}
\affiliation{Advanced Research Center for Nanolithography, Science Park 106, 1098 XG Amsterdam, the Netherlands}
\affiliation{Universiteit van Amsterdam, Science Park 904, 1098 XH Amsterdam, the Netherlands}

\begin{abstract}
We report on the detection of diffraction gratings buried below a stack of tens of 18 nm thick $\mathrm{SiO_2}$ and $\mathrm{Si_3N_4}$ layers and an optically opaque metal layer, using laser-induced, extremely-high frequency ultrasound. In our experiments, the shape and amplitude of a buried metal grating is encoded on the spatial phase of the reflected acoustic wave. This  grating-shaped acoustic echo from the buried grating is detected by diffraction of a delayed probe pulse. The shape and strength of the time-dependent diffraction signal can be accurately predicted using a 2D numerical model. Surprisingly, our numerical calculations show that the diffracted signal strength is not strongly influenced by the number of dielectric layers through which the acoustic wave has to propagate. Replacing the $\mathrm{SiO_2}$/$\mathrm{Si_3N_4}$ layer stack with a single layer having an equivalent time-averaged sound velocity and average density, has only a small effect on the shape and amplitude of the diffracted signal as a function of time. Our results show that laser-induced ultrasound is a promising technique for sub-surface nano-metrology applications.
\end{abstract}

\maketitle


\section{Introduction}

\begin{figure*}
\centering
\includegraphics[scale=0.65]{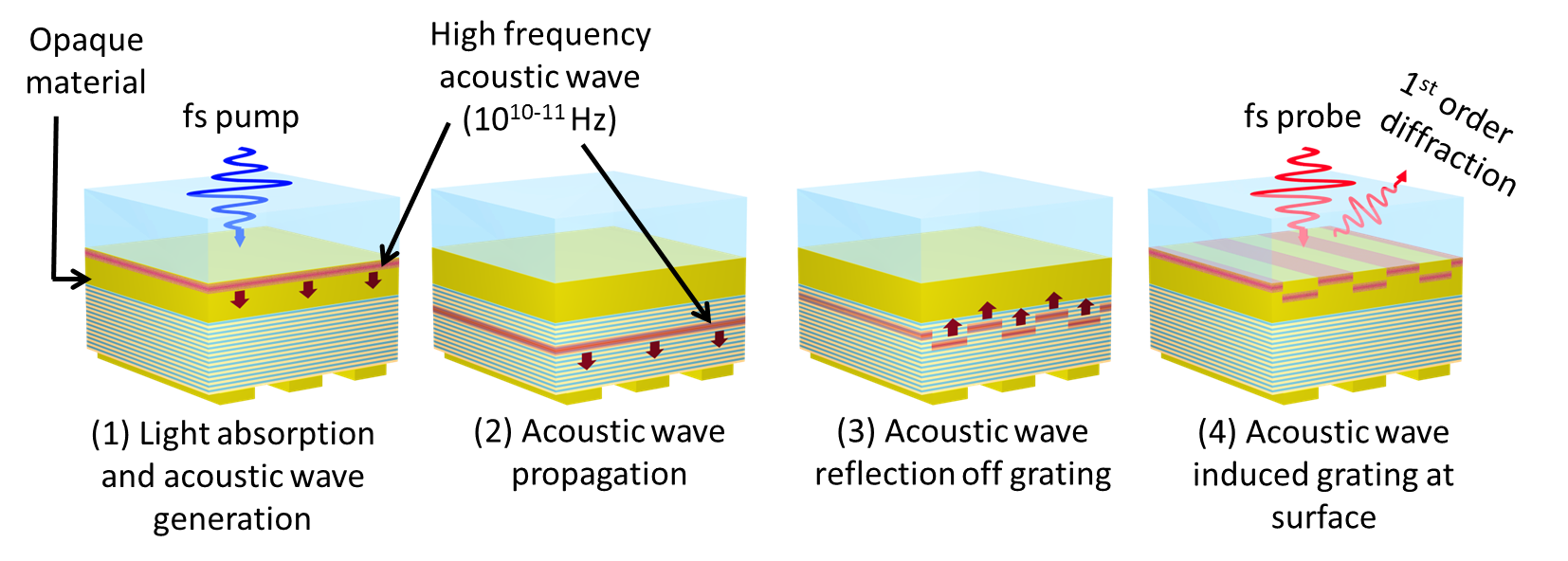}
\caption{Schematic explaining the technique. The femtosecond laser pulse is absorbed by the material at the substrate-material interface (1), launches an acoustic wave that propagates through different layers (2). The acoustic wave reflects off the buried grating and returns to the substrate-material interface as a grating-shaped acoustic wave (3).  The time-delayed femtosecond probe pulse diffracts off this interface grating, and the first-order diffraction signal is recorded (4).}
\label{intro_fig}
\end{figure*}

In semiconductor device manufacturing, techniques to detect micro- and nano-structures buried below the surface of deposited layers are extremely important \cite{metro1,metro2,metro3,metro4}. An example is the detection of so-called alignment gratings. Alignment gratings are gratings etched into Si wafers that are used to position wafers with sub-nanometer accuracy. This is done by illuminating the grating with a light source and by measuring the diffracted orders emerging from the grating. A small translation of the wafer in the direction of the grating wavevector has no effect on the diffraction efficiency. However, it does change the optical phase-difference between the $\mathrm{+n^{th}}$ and $\mathrm{-n^{th}}$  diffracted orders (with n=1,2,3,..). A change in the phase difference between, for example, the $\mathrm{+1^{st}}$ and $\mathrm{-1^{st}}$  order diffracted beams can accurately be detected by interfering the two beams. Measuring these changes makes it possible to align wafers with an accuracy of less than a nanometer \cite{den_boef}. Unfortunately, as device architectures become increasingly complex, alignment gratings can become buried below a large number of dielectric and/or metallic layers deposited during device fabrication. This poses a huge challenge for the detection of alignment gratings, in particular because some of these layers are barely transparent to light. Without sufficient diffracted light, wafer alignment and fabrication of nano-scale devices become very difficult.

Fortunately, layers that are opaque to light are often transparent to sound. It has been shown that femtosecond and picosecond laser pulses can be used to generate and detect sound waves with frequencies in the range of tens of gigahertz to several terahertz in solid opaque materials \cite{ac1,ac2,ac3,ac4,ac5,ac6,ac7,ac8,ac9,ac10,ac11,ac12,ac13,ac15,ac16,ac17,ac18,ac19,ac20,ac22,ac23,ac24,ac25,ac26,ac27,ac28,ac29,ac30,ac31,ac32}. These experiments were mostly performed on relatively simple systems consisting of one or a few layers and have led to a vastly improved understanding of laser-induced ultrasound. Therefore, using laser-induced ultrasound could be a novel and appealing approach to detect gratings buried underneath optically opaque layers. The challenge, however, is not only to detect gratings underneath opaque layers with these extremely-high-frequency sound waves. The challenge is also to do this through complex multi-layered systems that can be found in state-of-the-art semiconductor devices, such as in 3D NAND memory \cite{metro5}.

Here, we show how we can detect buried gratings underneath optically opaque layers, by measuring transient optical diffraction from ultrafast, laser-induced, extremely high-frequency acoustic copies of the grating. In our proof-of-principle experiment we first fabricate metal gratings on top of a single metal layer deposited on a glass substrate. When viewed from the glass side, the gratings are essentially invisible and can be considered ``buried''. We performed femtosecond pump-probe experiments with 400 nm wavelength pump and 800 nm wavelength probe from a Ti:Sapphire amplifier. As illustrated in Fig \ref{intro_fig}, an optical pump-pulse excites  the metal through the substrate and launches an acoustic wave which propagates through the opaque layer, then reflects off the peaks and valleys of the grating and thus acquires a spatially periodic phase. This wave returns to the glass-metal interface where it deforms the interface in a spatially periodic manner. This interface grating can be detected by diffraction of a time-delayed probe-pulse.  Our measurements show that on simple systems consisting of 10 nm amplitude buried gratings underneath thick gold (Au) or nickel (Ni) layers, diffraction is easily detected. A comparison with calculations shows that  grating-like deformation at the glass-metal interface induced by the acoustic echo, has an amplitude of several tens of picometers. Remarkably, we also observe diffraction at the glass-metal interface on more complex systems consisting of 10 nm amplitude gratings fabricated on top of  5 or 10 bilayers of thin silicon dioxide ($\mathrm{SiO_2}$) and silicon nitride ($\mathrm{Si_3N_4}$) layers on top of a Au or Ni  layer on glass. For the sample with 10 bilayers of $\mathrm{SiO_2}$ and $\mathrm{Si_3N_4}$ layers, after being generated, the acoustic wave has to travel through 42 layers in total before the acoustic echo reaches the glass-metal interface again where it is detected by diffraction of the optical probe pulse. Surprisingly, we find that the diffraction signal strength is not strongly influenced by the number of layers in the stack. This is confirmed by numerical calculations showing that replacing the $\mathrm{SiO_2}$/$\mathrm{Si_3N_4}$ stack with a single layer having an equivalent time-averaged sound velocity and average density, has only a small effect on the shape and amplitude of the diffracted signal as a function of time. The calculations demonstrate that the complex shape of the time-dependent diffracted signal is predominantly influenced by reflections of the acoustic wave at the glass-metal and the metal-stack interfaces. Our results show that buried gratings can be detected through optically opaque layers on complex, multi-layered samples, using laser-induced, extremely high-frequency ultrasound. This technique shows promise as a new, non-contact, all-optical grating detection- and imaging-modality for wafer alignment applications by using ultrasound to make an acoustic copy of the buried grating, while using conventional optical diffraction to read-out the copy when it reaches the surface. 

\section{Experimental Details}
\begin{figure*}[!]
\centering

\includegraphics[scale=0.7]{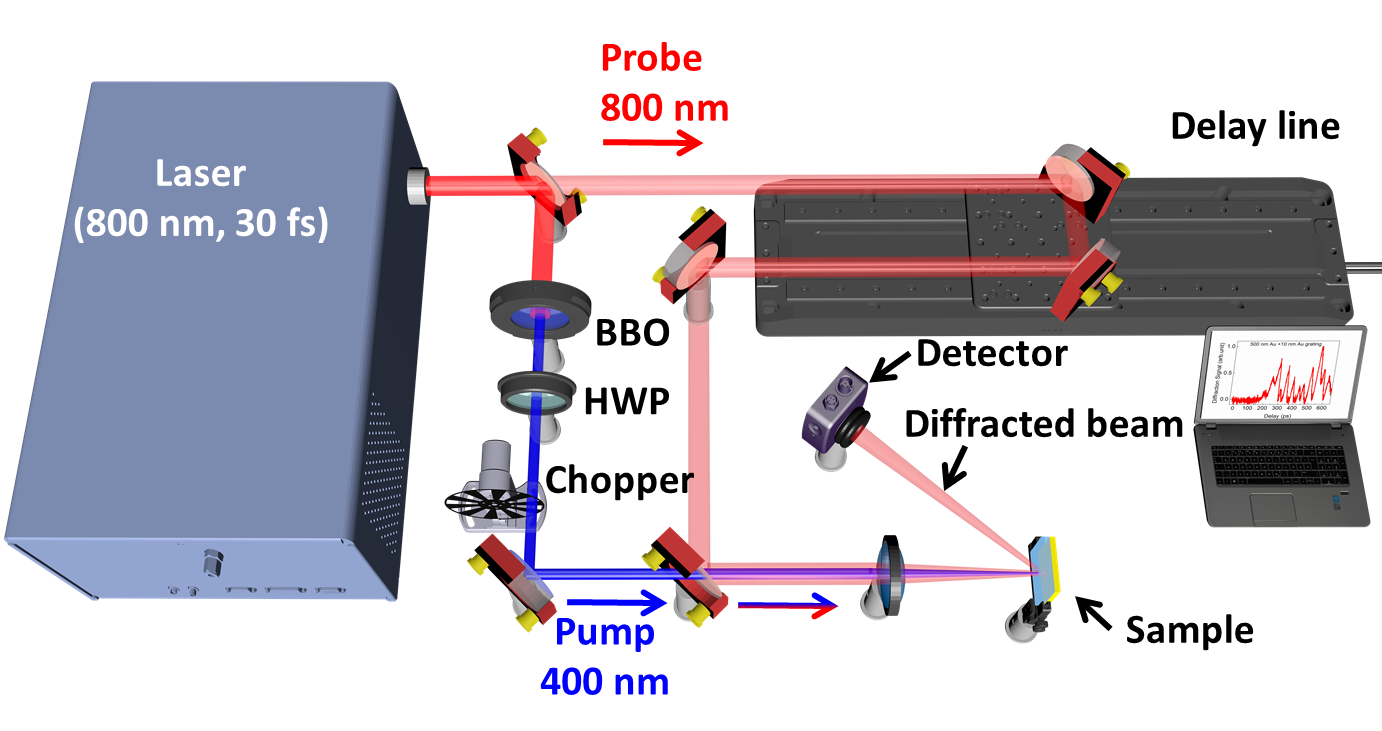}

\caption{Schematic of the experimental setup. 30 fs pulses with a wavelength of 800 nm are split into pump and probe beams. The pump is frequency-doubled in a BBO crystal, passes through a half-wave plate (HWP) and is focused onto the sample with a lens. The 800 nm probe beam passes through a variable optical delay line and is focused onto the sample on the same spot as the pump beam. An optical detector measures the first-order diffracted probe beam. }

\label{setup}
\end{figure*}
A schematic of the pump-probe setup used for the experiments is shown in Fig \ref{setup}. A Ti: Sapphire multi-pass amplifier  generates 30 fs pulses with a repetition rate of 1 kHz and with a wavelength centered at 800 nm. The output from the laser is split into two by a 95/5 beamsplitter. A 100 $\mathrm{\mu m}$ BBO crystal frequency-doubles the stronger beam to generate 400 nm pump pulses. The weaker part off the 800 nm beam is used as a probe. A half-wave plate (HWP) rotates the polarization of the 400 nm pump pulse by $\mathrm{90^{\circ}}$ so that both the pump and the probe are p-polarized. Both the pump and probe are weakly focused onto the sample such that the pump beam diameter is about 500 $\mathrm{\mu m}$ and pump pulse energy ranges from 6 $\mathrm{\mu J}$ to 8 $\mathrm{\mu J}$ depending on the sample. The probe beam diameter is 250 $\mathrm{\mu m}$ and the probe pulse energy was kept constant at 2 $\mathrm{\mu J}$. A silicon photo-detector is placed at the position where the first-order diffracted probe beam from the buried grating is expected. The signal recorded by the detector when the chopper blocks the pump beam is subtracted from the diffracted probe signal when the pump beam is transmitted by the chopper, and plotted as a function of the pump-probe delay. 

\subsection{Sample fabrication}

In principle, buried gratings can be made by chemically etching gratings in Si, followed by the deposition of dielectric and/or metallic layers. The resulting sample, however, then shows a strong surface topography which follows the topography of the buried grating even though it is not a true copy of it. When real semiconductor devices are manufactured, repeated steps involving deposition of layers followed by polishing are carried out, ultimately reducing or eliminating the residual surface topography. Making samples with zero surface topography turned out to be impossible using our clean-room facilities. To test whether laser induced ultrasonics  is capable of detecting buried gratings, we therefore opted to make samples by first depositing nominally flat dielectric/metallic layers on glass followed by the fabrication of a grating on top of this. By now performing pump-probe diffraction measurements from the glass side, the grating is invisible to both pump and probe and can be viewed as a buried grating. All the samples were prepared on  175 $\mathrm{\mu m}$ thick, chemically cleaned glass substrates. The Au and Ni layers were fabricated by physical vapor deposition, and the thickness was determined by a quartz crystal thickness monitor. The gratings on top of the metal/dielectric layers were fabricated by UV optical lithography. All gratings used in this research have a pitch of 6 $\mathrm{\mu m}$. The $\mathrm{SiO_2}$ and $\mathrm{Si_3N_4}$ layers were deposited by sputtering, using a silicon target in the presence of oxygen and nitrogen, respectively. To calibrate the thickness of the $\mathrm{SiO_2}$ and $\mathrm{Si_3N_4}$ layers, we performed linear spectroscopy measurements on single layers of $\mathrm{SiO_2}$, and $\mathrm{Si_3N_4}$ deposited on Si, under the same conditions.

\section{Numerical Simulation}
An advanced 2D numerical model which captures the generation, propagation and detection of high-frequency acoustic waves by ultrafast laser pulses is used to simulate the diffracted signal. The model consists of three main parts, (i) absorption of light and the subsequent generation of the acoustic wave, (ii) propagation of the acoustic wave, (iii) detection of the acoustic wave. The absorption of the femtosecond laser pulse and the subsequent heating and cooling of the electron gas inside a metal layer is described by the well-known Two Temperature Model (TTM)\cite{ttm1,ed1,ed2,ed3,ed4,ed5}. The heating of the lattice calculated from the TTM sets up an isotropic thermal stress, which leads to the generation of the high-frequency acoustic wave. The equation of motion for an isotropic, linear elastic wave is used to describe the propagation of the acoustic wave inside the metal and dielectric layers. Finally, by propagating the complex electric field of the probe pulse  after the optical excitation by the pump pulse, through the substrate and into the metal, we can calculate the diffraction efficiency as a function of time delay. The model calculates the first-order diffracted signal by  accounting for the spatially periodic changes in refractive index due to the thermo-optic effect and the strain-optic effect, and the spatially periodic displacement of the glass-metal interface\cite{hao}.

\section{Results and discussion}
\subsection{Pump-probe measurements}

\begin{figure*}
\hspace{0.5cm}
\includegraphics[width=0.4\textwidth]{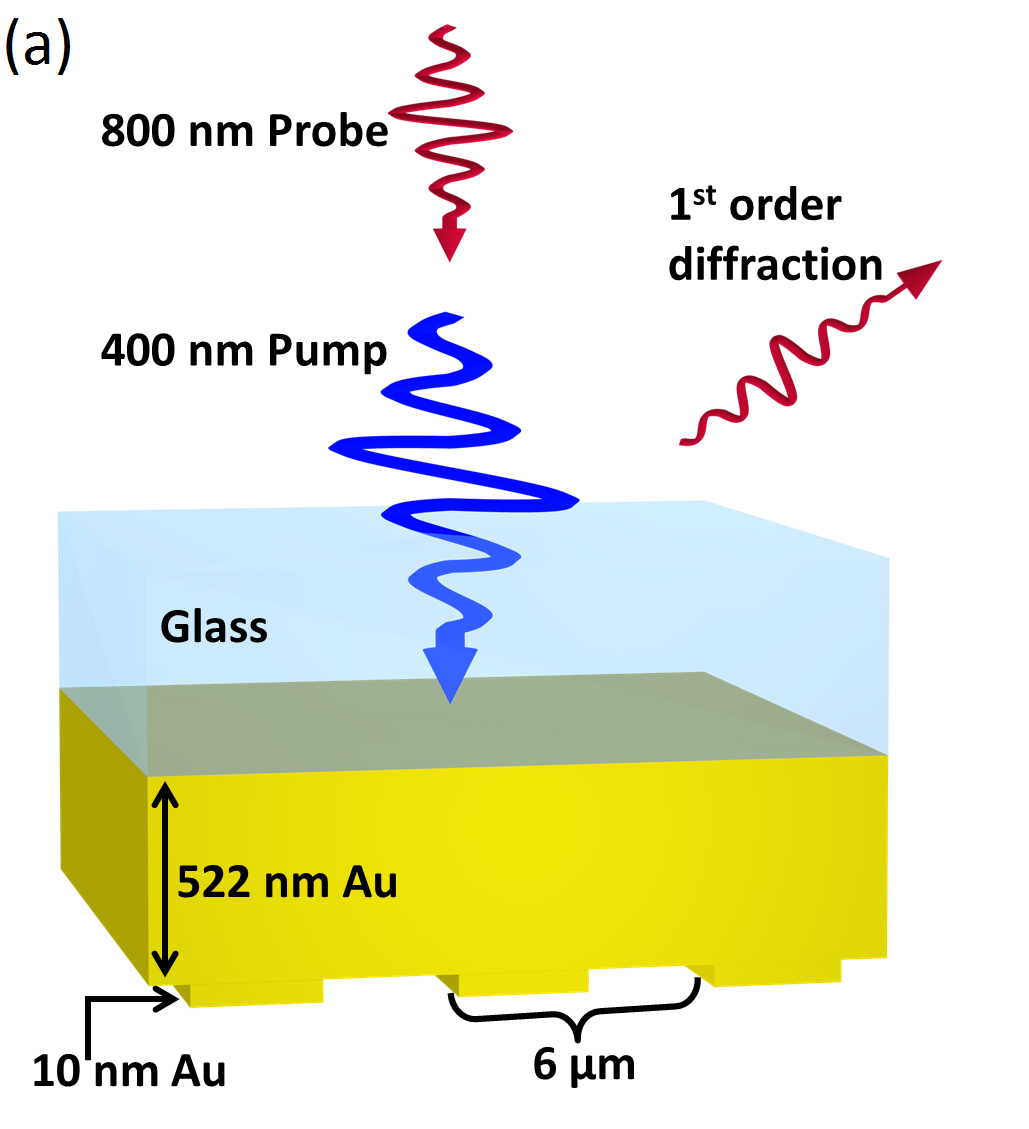}
\hspace{1.0cm}
\includegraphics[width=0.4\textwidth]{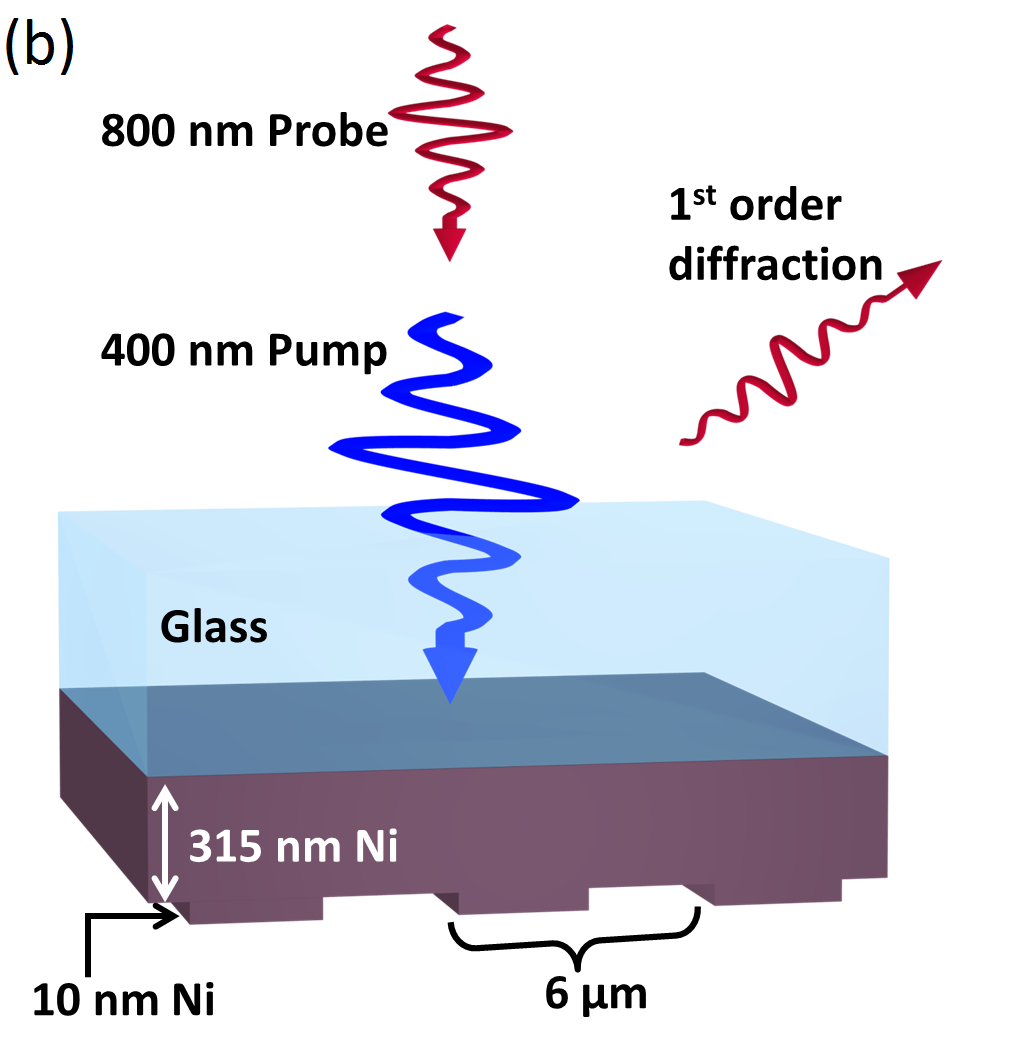}
\vspace{1.75cm}
\hspace{0cm}
\includegraphics[width=0.4\textwidth]{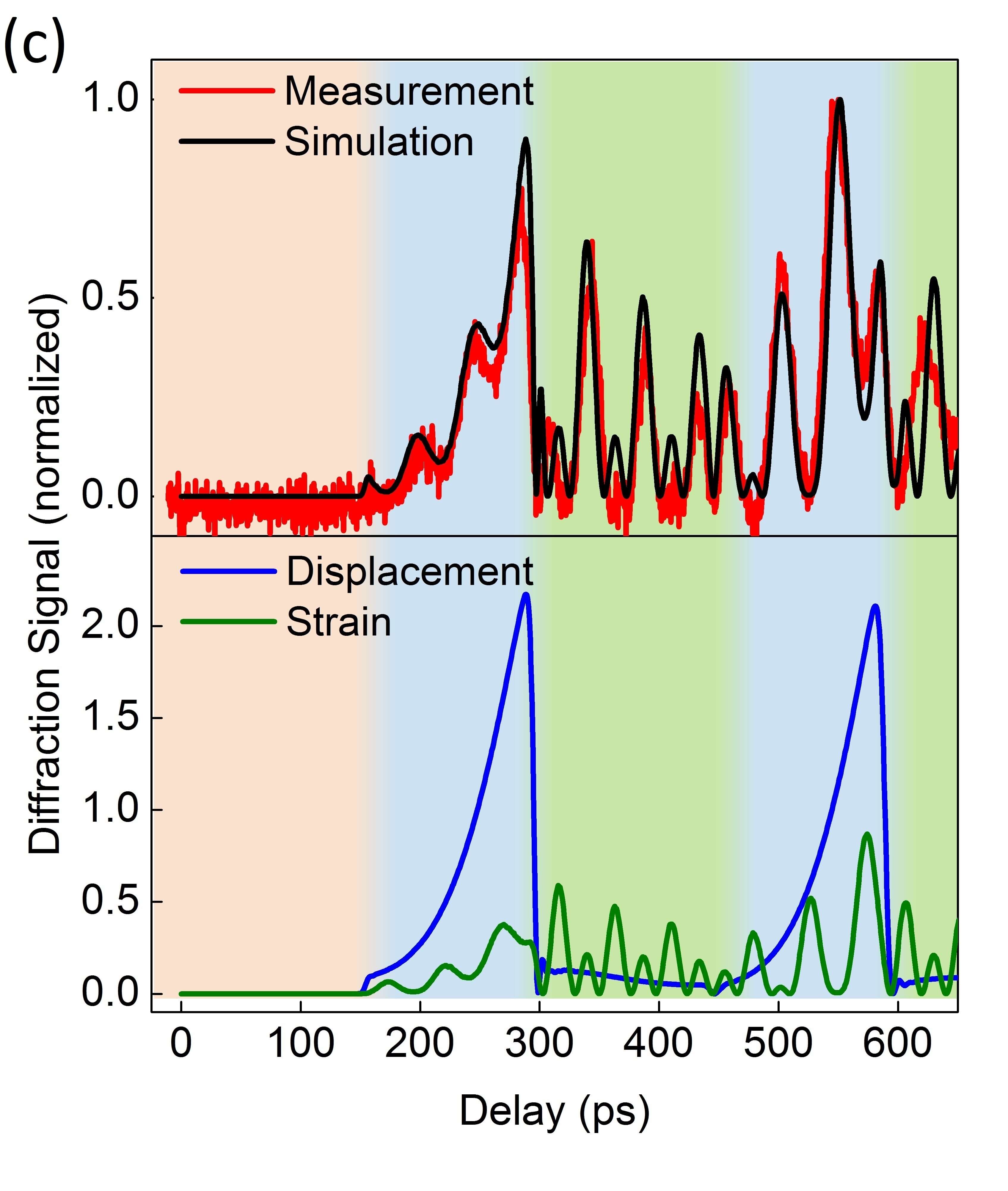}
\hspace{0cm}
\includegraphics[width=0.4\textwidth]{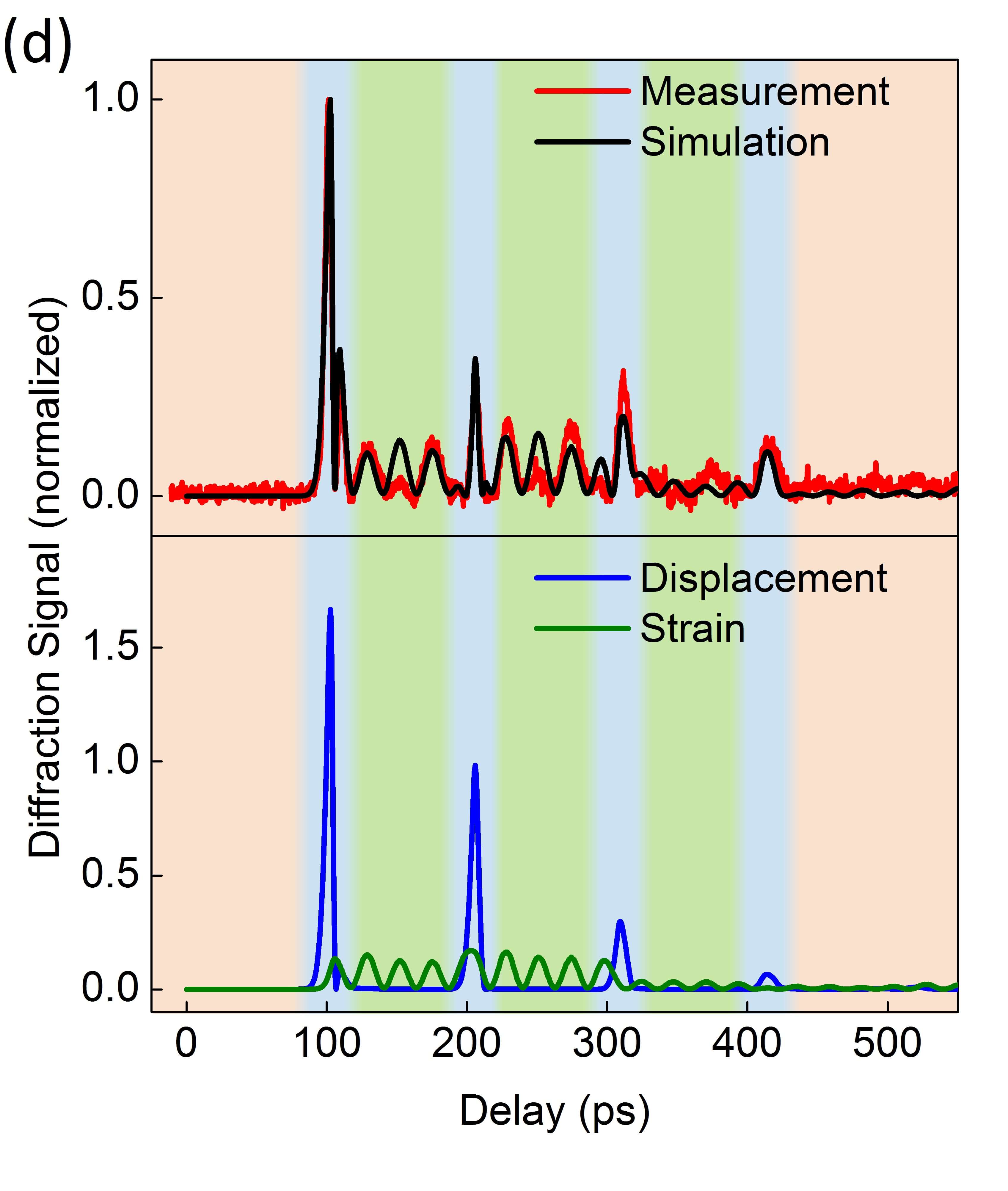}

\caption{Schematics of the beam/sample geometry for: (a) the 10 nm amplitude Au grating on a 522 nm Au layer on glass and (b) 10 nm amplitude Ni grating on a 315 nm Ni on glass. Both gratings have a 50 $\%$ duty-cycle. The pump pulse has a wavelength of  400 nm  and the probe pulse has a wavelength of 800 nm. (c) Upper  panel: The experimentally measured (red) and numerically simulated (black) diffracted probe signal vs. pump-probe delay for the Au on Au grating sample. Bottom panel: Calculated probe diffraction signal vs. pump-probe delay taking only the displacement of the glass-Au interface into account (blue line), or taking only the propagating strain pulse in the glass substrate into account (green line). (d) Upper panel: The experimentally measured (red) and numerically simulated (black) diffracted probe signal vs. pump-probe delay for the Ni on Ni grating sample. Bottom panel: Calculated probe diffraction signal vs. pump-probe delay taking only the displacement of the glass-Ni interface into account (blue line), and taking only the propagating strain pulse in the glass substrate into account (green line).}
\label{Au_Ni_plots}
\end{figure*}

To get an estimate of the nature and strength of the diffracted signal from the acoustic echo of the buried grating, we perform femtosecond, time-resolved experiments on relatively simple systems, consisting of (i) a 10 nm amplitude 50 $\%$ duty cycle Au grating on a 522 nm flat Au layer deposited on glass (Fig \ref{Au_Ni_plots}(a)) and, (ii) a 10 nm amplitude, 50 $\%$ duty cycle Ni grating on a 315 nm flat Ni layer deposited on glass (Fig \ref{Au_Ni_plots}(b)). Au is rarely used in the semiconductor manufacturing industry, and our choice for Au as the grating/layer material is exclusively motivated by the fact that Au is one of the most well-studied materials.  However, it has a relatively small electron-phonon coupling constant compared to that of Ni, and we expect this to have a significant impact on the shape of the resulting acoustic signals\cite{ed2,aug1}. All pump-probe measurements discussed in this paper were performed from the substrate side. As the gratings are fabricated at the metal-air side and are thus optically hidden from the pump and probe pulses (see Fig \ref{Au_Ni_plots}(a)), the gratings can be considered ``buried''. 

In Fig \ref{Au_Ni_plots}(c), we plot the measured first-order diffracted signal (red curve) as a function of the pump-probe delay for the sample consisting of a 10 nm amplitude Au grating on a 522 nm flat Au layer on glass. The graph shows that for the first 150$\mathrm{\pm2}$ ps after optical excitation, there is no diffracted signal. After this, a diffraction signal emerges that slowly rises to a maximum at a delay of around 280$\mathrm{\pm2}$ ps. A second maximum can be seen around delay values of about 560 ps. Superimposed on these is a more rapidly oscillating signal with a period of 47$\mathrm{\pm2}$ ps. Note that the signature of this more rapid oscillation is already present on the rising edge of the first main diffraction peak but also on the falling edge of the second main diffraction peak.  

We want to point out that the very observation of diffraction after 150 ps implies the presence of a grating at the glass-Au interface. This grating is caused by the returning acoustic wave whose wavefront has the topography of the buried grating imprinted on it through a spatially periodic delay in the acoustic arrival time, or phase. In principle, the nature of this grating is a displacement of the atoms at the glass-Au interface. For other materials, spatially periodic changes in the optical constants of the materials, caused by changes in material density induced by the sound wave, can also contribute. As we will show later, for Au, our calculations indicate that displacement of the atoms at the glass-Au interface leading to a grating with an amplitude of up to  110 pm at the glass-Au interface is the dominant cause of diffraction. Regardless, our measurements show that we can observe the presence of an optically hidden, 10 nm amplitude, grating through diffraction from an acoustic ``copy'' of the grating.

To properly understand these measurements we first focus on the acoustic wave generation mechanism in Au. The 400 nm pump pulse is absorbed within the optical penetration depth of about 16 nm in Au, creating a hot electron gas \cite{aupd,ed2,stephen}.  The electron-gas energy rapidly diffuses deeper into the thick Au layer to a maximum depth of several hundred nanometers. Such a large diffusion depth is possible because Au has a relatively small electron-phonon coupling strength \cite{aug1,aug2}. On a time scale of a few picoseconds, the electron gas cools by heating the lattice. The highest lattice temperatures are found near the glass-Au interface, where the light is absorbed. The rapid heating of the lattice leads to a rapid expansion of the lattice which launches a longitudinal acoustic pulse with a spatial extent corresponding to the electron energy diffusion depth. The slow, early rise of the diffraction signal versus time is due to the arrival of part of the acoustic wave that has been generated deep inside the Au layer and thus is the earliest to reflect off the buried grating.  The diffraction signal continues to slowly increase up to 280$\mathrm{\pm2}$ ps at which time the part of the acoustic wave that was generated near the glass-Au interface returns to the interface again after reflecting off the buried grating. A second, broad diffraction signal is seen around 550-580 ps which is the arrival of the same acoustic wave after the second round trip.

The rapid, lower amplitude oscillation superimposed on the signal is surprising, considering that its period doesn't match the round trip propagation time in 522 nm Au.  We attribute this oscillation to Brillouin scattering from the part of the acoustic wave that has entered the glass and continues to propagate \cite{bs1,bs2,bs3,bs4,bs5}.  The wavefront of this acoustic wave is still spatially periodic, modifies the refractive index of the glass, and gives rise to probe pulse diffraction. Normally, the presence of such a quasi-static grating in glass would give rise to a constant (in time) diffracted signal. Here, however, the same probe pulse that diffracts off this grating also reflects from the glass-Au interface and then diffracts off the grating in the same direction again \cite{bs3}. The extra optical phase acquired by the propagation of the optical pulse before it diffracts a second time leads to interference between the electric fields of the diffracted beams. Whether the interference is constructive or destructive depends on the distance that the optical pulse has propagated before diffracting again. This in turn depends on the distance that the acoustic wave in the glass has travelled from the glass-Au interface and is thus a periodic function of time. It can be shown \cite{bs1,bs2,bs3,bs4,bs5} that the period of this oscillation is given by $ T= \lambda / ( 2 n v cos(\theta))$.  Here  $\lambda$ is the probe wavelength, n is the refractive index of glass, v is the sound velocity in glass, and $\theta$ is the angle of incidence. In our experiments we have $ \lambda$= 800 nm, $n$=1.5, $v$= 5,700 m/s and $cos(\theta) \approx 1$,  giving an oscillation period of 46.7 ps which matches  the period observed in the measurements. The Brillouin scattering in glass dominates the time-dependent diffraction signal we observe between 300 ps and 450 ps. The diffraction peak at 283$\mathrm{\pm2}$ ps is from the first round trip of the acoustic wave, and the peak at 580$\mathrm{\pm2}$ is from the second round trip of the acoustic wave in 522 nm Au. As the acoustic wave travels back and forth in the Au layer, every time it reaches the glass-Au interface, part of the wave will be transmitted into the glass. The diffraction signal we observe after 450 ps therefore is a coherent sum of diffracted signal from multiple contributions, namely (i) light diffracting from the grating-shaped acoustic echo in Au when it is close to the  glass-Au interface after the second round trip, (ii) light diffracting due to Brillouin scattering in glass from the  acoustic wave grating that entered the glass after the first round trip and, (iii)  light diffracting due to  Brillouin scattering in glass from the acoustic wave that entered the glass after the second round trip. Interference between the light beams diffracted off these gratings makes it difficult to predict what the temporal shape of the diffracted signal will look like. As we will show below, we therefore resorted to 2D numerical calculations of the generation, propagation, diffraction, and optical detection of the acoustic waves, which take into account all the aforementioned contributions. 

Finally, we note that shortly after optical excitation with a pump beam, the abrupt surface expansion of the metal layer also generates an acoustic wave which propagates into the glass. However, this acoustic wave does not result in the diffraction of the probe pulse, as its wavefront has not reflected off the grating and is, thus, ``flat''. 

Now that we understand the factors that contribute to the diffraction signal, we performed numerical simulations to show how these factors affect the shape and time evolution of the diffraction signal. The simulations are based on a numerical model we developed which solves a set of equations by 2D finite-difference  time-domain method. Those equations describe the generation, propagation and optical detection of the acoustic waves. The model includes diffraction cased by the displacement of the surface at the glass-Au interface, diffraction by possible changes in the complex refractive index in the metal, and diffraction of light by the refractive index grating formed by the acoustic waves in the glass. The model is briefly explained in the earlier section and  details of the model can be found in elsewhere\cite{hao}. In Fig \ref{Au_Ni_plots}(c) (upper panel), we plot the numerically simulated diffraction signal as a function of the pump-probe delay for the 10 nm amplitude grating on 522 nm Au on glass (black curve). The simulation is in good agreement with the experimentally measured diffraction signal, as it predicts both the position and the relative intensities of the diffraction peaks. The bottom panel in the same figure shows the contributions to the calculated diffraction signal due to the (i) displacement of the atoms near the glass-metal interface caused by the acoustic wave propagating in the Au layer (blue curve), (ii) changes in the optical constants of the glass due to the strain-optic effect, or Brillouin scattering, in the glass substrate (green curve). Note that the full calculations (black curve) cannot simply be obtained by adding the curves calculated for the strain contribution (green curve), to that for the curve calculated for surface atom displacement only (blue curve). The reason for this is that the phases of the diffracted light scattered by the two types of gratings have to be taken into account as well.  This can give rise to destructive interference and thus a net diffracted signal that is smaller than the simple sum of the intensities of the diffracted beams obtained for each grating separately. The model calculates the diffracted optical field, thus those coherent additions are fully taken into account.

To study the effect that the electron-phonon coupling strength has on the shape of the acoustic signal reflected off the buried grating, we also performed measurement on a 10 nm amplitude Ni grating fabricated on a flat 315 nm thick Ni layer on glass (Fig  \ref{Au_Ni_plots}(b)). The electron-phonon coupling constant in Ni is about 12 times larger than that of Au (see Table \ref{optical_ppt}. In Fig  \ref{Au_Ni_plots}(d) we plot the measured first-order diffracted signal as a function of the pump-probe delay for this sample (red curve). The diffraction signal stays zero for the first 100 ps after optical excitation, after which it rises sharply within 5 ps and quickly drops to zero, in contrast to the results shown for Au where the signal increases more slowly. In Ni, the hot electron gas formed after optical excitation cools much more rapidly and heats the lattice before the hot electron gas energy can diffuse much deeper into the layer. This rapid heating within a limited depth of $ \approx$ 75 nm results in the generation of a much higher frequency longitudinal acoustic wave. This acoustic wave is much more localized in the propagation direction than in the case of Au and thus gives rise to more narrow peaks in the diffraction signals. The diffraction peaks we observe at 102$\mathrm{\pm2}$ ps, 205$\mathrm{\pm2}$ ps, 310$\mathrm{\pm2}$ ps, and 415$\mathrm{\pm2}$ ps are due to diffraction of the probe pulse after the $\mathrm{1^{st}}$, the $\mathrm{2^{nd}}$, the $\mathrm{3^{rd}}$ and the $\mathrm{4^{th}}$ round trip of the acoustic echo that originated at the glass-Ni interface, respectively. 

One noticeable feature in these measurements is the very sharp decrease of the diffraction signal at 102 ps, immediately after the first main diffraction peak. The sharp dip can be understood from the shape of the wavefront of the acoustic wave after it reflects off the buried grating. The part of the wave that reflects off the valleys of the grating takes $\approx$ 3.5 ps longer to return to the glass-Ni interface than the part that reflects off the peaks of the grating. In fact, we can view the two parts as two separate time-delayed gratings, shifted by half-a-grating-period, equivalent to a  $\pi$ phase shift,  in the direction parallel to the grating vector.  When both gratings are close to the interface, within the region that corresponds to the optical penetration depth of the probe pulse, the probe can diffract off both gratings simultaneously. However, a $\pi$ phase difference between the two gratings implies a $\pi$ phase difference between the electric fields diffracted off the two gratings. Therefore, soon after the arrival of the first acoustic grating,  the second acoustic grating arrives, and the field diffracted off this grating destructively interferes with that diffracted off the first. This leads to a very sharp dip in the diffracted intensity immediately after the first diffraction peak. 

Between 110 ps and 190 ps, we observe diffraction peaks caused by  Brillouin scattering in the glass substrate. In contrast to the measurement on the Au sample described above, Brillouin scattering oscillation is not observed on the rising edge of the diffracted signal from the first acoustic echo in the Ni layer. This is because the rise time of the acoustic wave amplitude is simply too short. By the time the acoustic wave arrives near the glass-Ni interface and is partially transmitted by it, the acoustic wave in the glass has not propagated far enough to observe the oscillation. This is different for the Au case where the acoustic wave is much longer. After about 115 ps, the acoustic echo is no longer present at the glass-Ni interface, and the small diffraction peaks at 130 ps and 176 ps are only due to Brillouin scattering in the glass. The separation between these peaks is 46 ps which is exactly the Brillouin oscillation period in glass for the probe wavelength. Similarly, after the second round trip of the acoustic wave in Ni at 205 ps, we observe small diffraction peaks separated by 46 ps at 230$\mathrm{\pm2}$ ps and 276$\mathrm{\pm2}$ ps which are also due to the Brillouin scattering. The short duration of the acoustic wave in Ni enables us to more easily separate the Brillouin scattering effect in the glass from that of the acoustic wave in the metal.  We observe that the width (FWHM) of the diffraction peaks due to the  acoustic wave  in Ni at 102 ps, 205 ps, 310 ps, and 415 ps, is 5$\mathrm{\pm1}$ ps, 7$\mathrm{\pm1}$ ps, 9$\mathrm{\pm1}$ ps, and 13$\mathrm{\pm1}$ ps respectively. This gradual increase in the FWHM of the diffraction signal after each round trip is suggestive of acoustic dispersion of the acoustic wave. However, the presence of the oscillations caused by Brillouin scattering make it difficult to ascertain whether this is the only factor contributing to the increase in the FWHM. No measurable diffraction by the acoustic wave propagating in Ni is observed after 450 ps. The main reason for this is that the acoustic wave undergoes damping as it propagates inside the Ni layer and because of acoustic reflection losses at the glass-Ni interface upon every round trip. 

In Fig \ref{Au_Ni_plots}(d)(upper panel), we plot the numerically calculated diffraction signal (black curve) along with the measured diffraction signal for the Ni sample.  In Fig \ref{Au_Ni_plots}(d) (lower panel), we plot the diffraction signals due to (i) the  displacement of the atoms near the glass-Ni interface (blue curve) and (ii)  Brillouin scattering in the glass substrate (green curve).  Here too, the calculated diffraction signal  that takes both surface displacement and Brillouin scattering into account (black curve), is very similar to the measured one. These calculations allow us to unambiguously attribute the largest diffraction peaks at 102 ps, 205 ps, 310 ps and 415 ps to scattering off the spatially periodic displacement of the atoms at the glass-Ni interface after each round trip of the acoustic wave inside the Ni layer. The lower amplitude diffraction peaks observed in between are clearly caused by Brillouin scattering in the glass. Note that the calculations also show hints of broadening of the diffracted peaks when only the interface displacement is taken into account. This again is suggestive of peak broadening by acoustic wave dispersion.

\subsection{Complex multilayer samples}
\begin{figure*}
\hspace{0.5cm}
\includegraphics[width=0.4\textwidth]{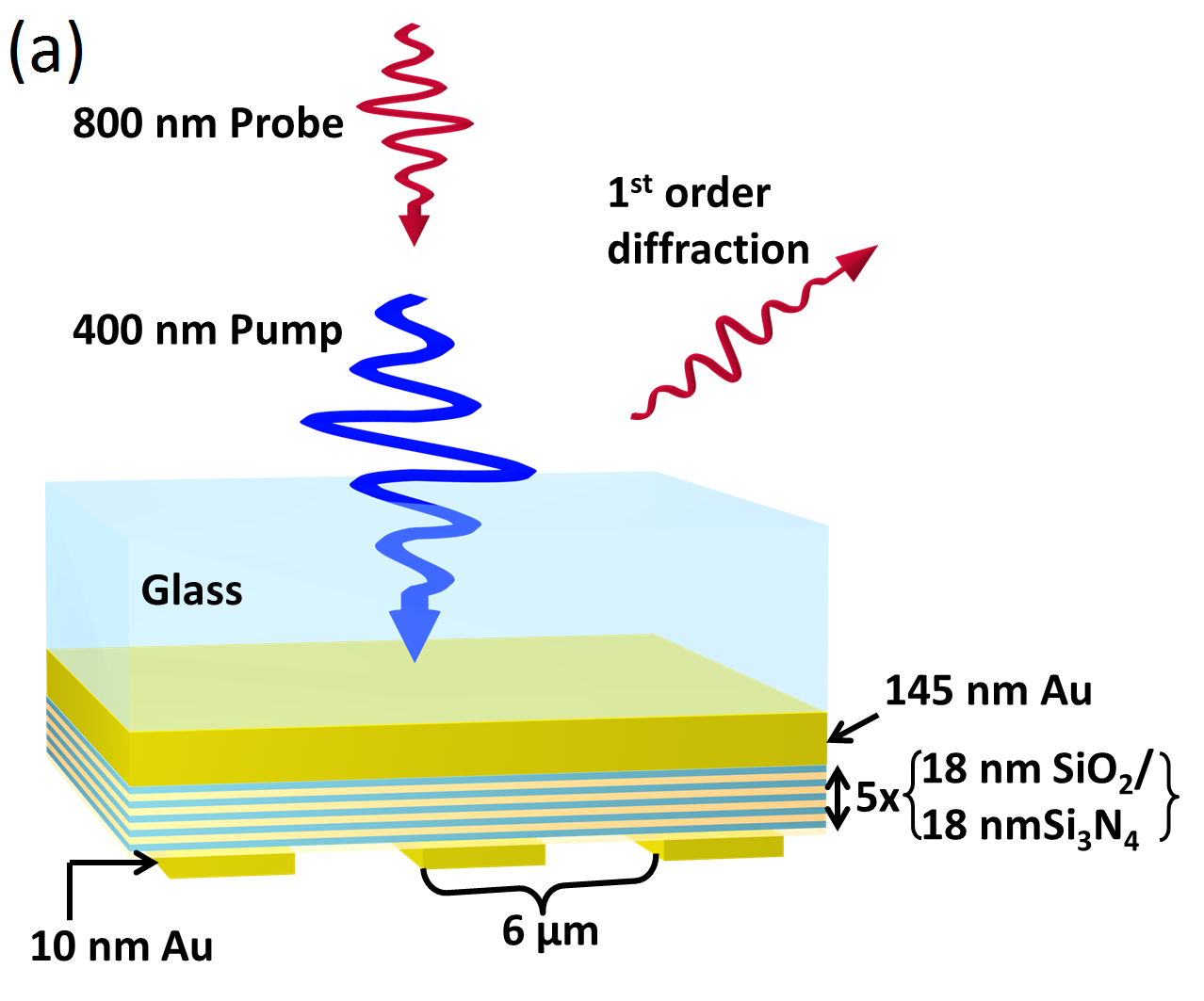}
\hspace{1.0cm}
\includegraphics[width=0.4\textwidth]{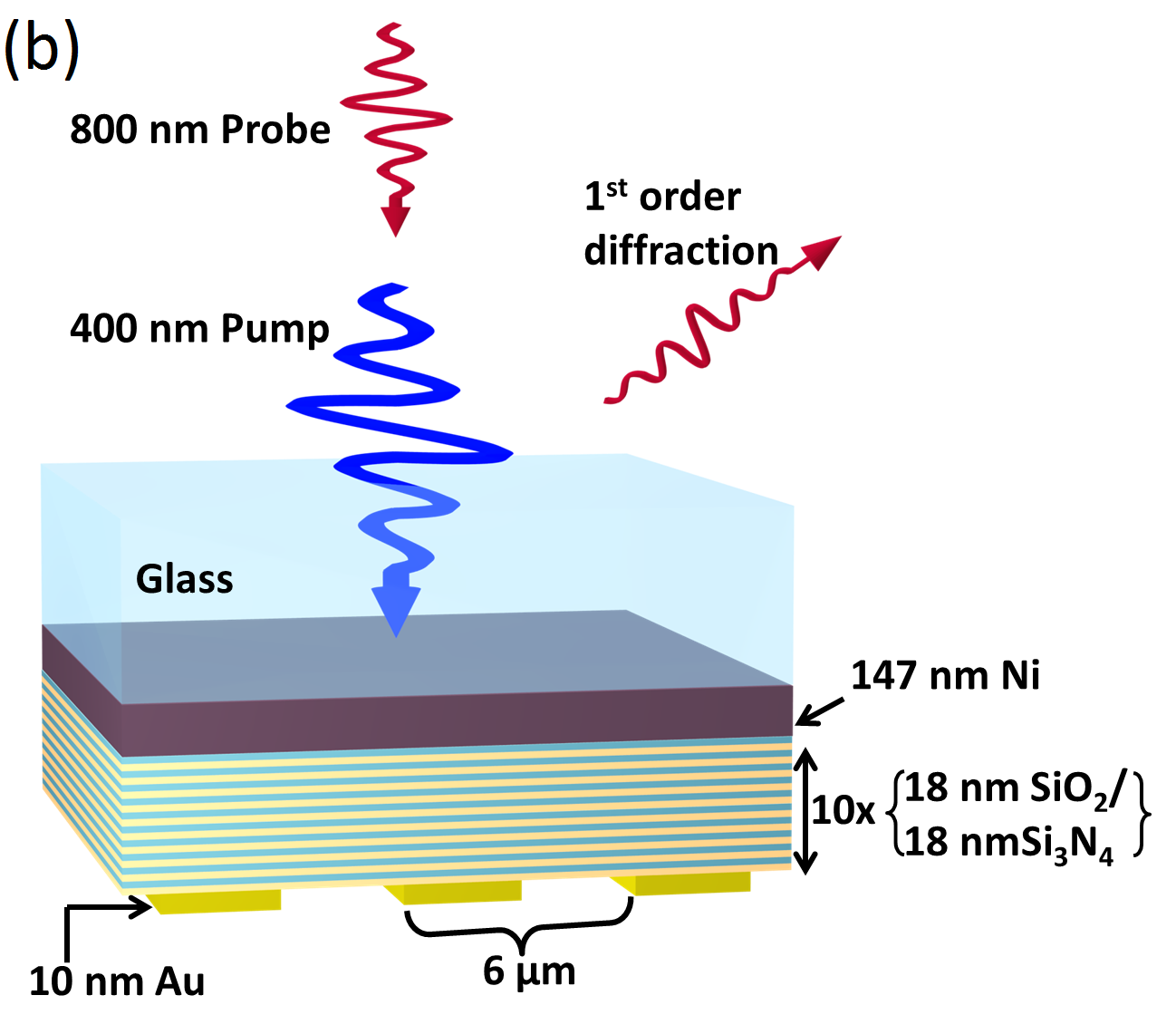}
\vspace{1.75cm}
\hspace{0cm}
\includegraphics[width=0.4\textwidth]{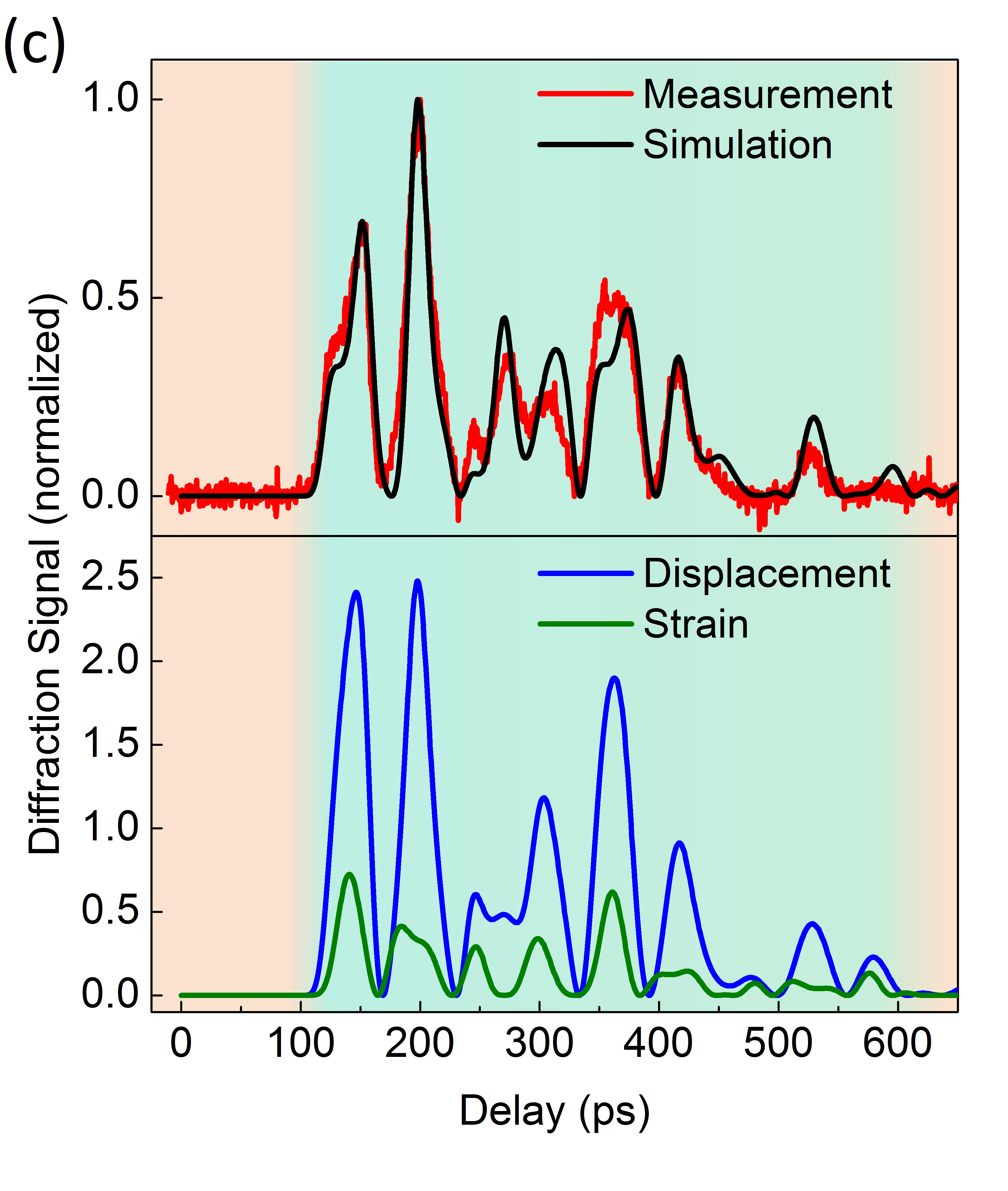}
\hspace{0cm}
\includegraphics[width=0.4\textwidth]{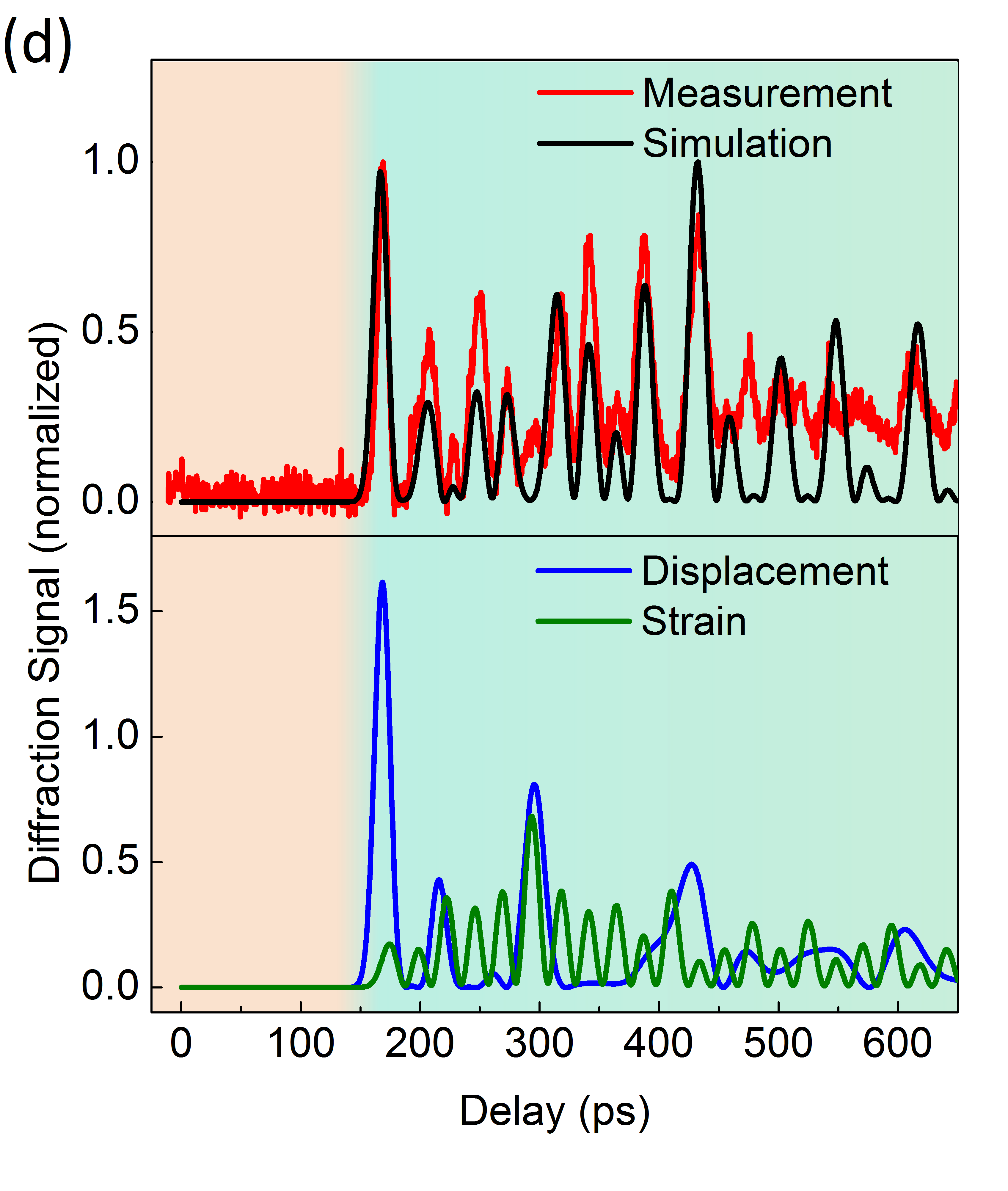}

\caption{ Schematics of the beam/sample geometry for: (a) the 10 nm amplitude Au grating on 5 pairs of alternating layers of 18 nm $\mathrm{SiO_2}$ and 18 nm $\mathrm{Si_3N_4}$ fabricated on a 145 nm thick Au layer on glass (``Au-multilayer''), (b) the 10 nm amplitude Au grating on 10 pairs of alternating layers of 18 nm $\mathrm{SiO_2}$ and 18 nm $\mathrm{Si_3N_4}$ , fabricated on top of a 147 nm thick Ni layer on glass (``Ni-multilayer'').  (c) Upper panel: The experimentally measured (red) and numerically simulated (black) diffracted probe signal vs. pump-probe delay for the ``Au-multilayer'' sample. Bottom panel: Calculated probe diffraction signal vs. pump-probe delay taking only the displacement of  the glass-Au interface into account (blue line), and taking only the propagating strain pulse in the glass substrate into account (green line). (d) Upper panel: The experimentally measured (red) and numerically simulated (black) diffracted probe signal vs. pump-probe delay for the ``Ni-multilayer'' sample. Bottom panel: Calculated probe diffraction signal vs. pump-probe delay taking only the displacement of the glass-Ni interface into account (blue line), and taking only the propagating strain pulse in the glass substrate into account (green line).}
\label{3D_plots}
\end{figure*}

Now that we understand the experiments performed on buried gratings deposited on single Au and Ni layers, we next performed measurements on more realistic and, thus, more complex samples.  In real-life semiconductor device manufacturing, alignment gratings are often etched in narrow scribe lanes between individual chips on a wafer. No devices are fabricated in these lanes, but the gratings get covered by the deposited opaque materials nonetheless. For example, manufacturing 3D NAND memory chips requires detection of alignment grating through thick opaque metal/dielectric layer and many alternating layers of oxide and nitride\cite{metro5}. The samples we fabricated consist of layers of materials that partially mimic the materials/structure used in the fabrication of 3D NAND memory. However, for comparison with the results shown for single layers, we used Au and Ni as the grating material and as the layer in which the sound waves are generated.

For the first sample, which will henceforth be referred to as the ``Au-multilayer'' sample, we evaporated 145 nm of Au on glass, on top of which we deposited five alternating pairs of 18 nm thick  $\mathrm{SiO_2}$ and 18 nm thick  $\mathrm{Si_3N_4}$. On top of this stack we fabricated a 10 nm amplitude Au grating, as shown in Fig \ref{3D_plots} (a). For the second sample, which will henceforth be referred to as the ``Ni-multilayer'' sample, we first deposited a 147 nm thick layer of Ni followed by the deposition of ten pairs of 18 nm thick $\mathrm{SiO_2}$/18 nm thick $\mathrm{Si_3N_4}$ layers. On top of this, a 10 nm amplitude  Au grating was fabricated. In this case, the $\mathrm{SiO_2}$/$\mathrm{Si_3N_4}$ stack consists of 20 layers in total and is thus twice as thick as the stack in the first sample. This sample is schematically shown in 2(b). The thickness of the $\mathrm{SiO_2}$ and $\mathrm{Si_3N_4}$ layers was calibrated by linear spectroscopy measurements. 

In Fig \ref{3D_plots}(c)  we show the measured diffraction signal as a function of the pump-probe delay (red curve), for the ``Au-multilayer'' sample. The diffraction signal remains zero for about 100 ps, after which the first diffraction peak is observed,  followed by a quasi-periodic oscillating signal. We emphasize that the time-dependent diffraction signal we observe in our measurements is proof that we detect the presence of a ``buried'' grating, by measuring optical diffraction off an acoustic copy of the grating near the glass-metal interface.  This means that, in spite of the many interfaces encountered by the propagating acoustic wave, a well-defined acoustic copy of the buried grating can still be  detected near the glass-Au interface. Remarkably, the energy of the acoustic wave has not completely dissipated as it propagates through these layers. Similar to the measurements on the grating  fabricated on a single layer of Au, the individual peaks are fairly broad. Again, this is caused by the relative homogeneous heating of the Au layer by rapid diffusion of the electron gas energy into the metal layer. As in the previous measurements, the signal we measure is the coherent sum of the optical fields diffracted off the grating-shaped acoustic wave at the Au-glass interface and off the grating-shaped acoustic wave in the glass. 

To better understand the measurement, we performed a numerical simulation of the diffracted signal versus time delay, which is shown in Fig \ref{3D_plots}(c), upper panel (black curve). We see a remarkable agreement between the measured and the simulated curves for this complex multilayer  sample, indicating that the model contains all the physics necessary to predict the salient features of our measurements. In the bottom panel we plot the calculated diffracted field taking only the displacement of the glass-Au interface  by the acoustic wave  into account (blue curve) and the diffracted field calculated taking only the changes in the optical constants of the glass due to the strain-optic effect, or Brillouin scattering, into account (green curve). For the ``Au-multilayer'' sample, the first peak for the displacement contribution (blue curve) is at 146 ps, which corresponds to the acoustic round trip time for the whole stack of dielectric layers.  

In Fig \ref{3D_plots}(d)  we plot the measured diffraction signal as a function of the pump-probe delay (red curve), for the ``Ni-multilayer'' sample. The diffraction signal remains zero for about 155 ps,  then begins to increase until a first maximum is observed at  169$\mathrm{\pm2}$ ps. This diffraction signal is due to the acoustic wave returning to the glass-Ni interface after one round trip through the whole stack of layers. This arrival time  matches the expected propagation time through all the layers (see Table \ref{acoustic_ppt}).  The diffraction signal then drops to zero and quasi-periodic oscillations are observed. Note that here too the diffraction peaks are ``sharper'' than for the ``Au-multilayer'' sample, in a manner similar to what is observed for the single layer samples. Again, this can be explained by the larger electron-phonon coupling constant of Ni, which leads to shorter acoustic waves. The signal we observe after 169 ps has contributions from the interference of optical fields diffracted off  acoustic waves near the glass-Ni interface, and off acoustic wave transmitted into the glass substrate. To better understand the measurement, we performed a numerical simulation of the diffracted signal, which is shown in Fig \ref{3D_plots}(d), (black curve). The position of the peaks in the simulated curve agrees well with the position of the peaks in the measurement. The amplitudes are seen to match less well. The blue curve in the bottom panel in the  figure shows the calculated diffracted field taking only the displacement of the glass-Ni interface into account, and the green curve shows the diffracted field calculated taking only the Brillouin scattering in glass into account. The first  peak seen in the calculation of the diffracted signal caused by only the displacement, at 169 ps, is due to the return of the first acoustic wave reflected off the buried grating after propagating through all the layers. This grating-shaped acoustic wave, now at the glass-Ni interface, undergoes another reflection inside  the 147 nm Ni layer before it returns to the glass-Ni interface again where it gives rise to the second peak at 215 ps.  The periodic oscillations due to  Brillouin scattering in the glass substrate  can be seen more clearly in  the ``Ni-multilayer'' sample as compared to the ``Au-multilayer'' sample, because of the shorter length of the acoustic wave.  Our measurement on the ``Ni-multilayer'' sample demonstrates that we can detect the acoustic wave even after it has propagated back and forth through 20 dielectric, so through 40 layers in total.

\subsection{Effective acoustic properties of the bilayer dielectric stack}

\begin{figure}

\includegraphics[width=0.45\textwidth]{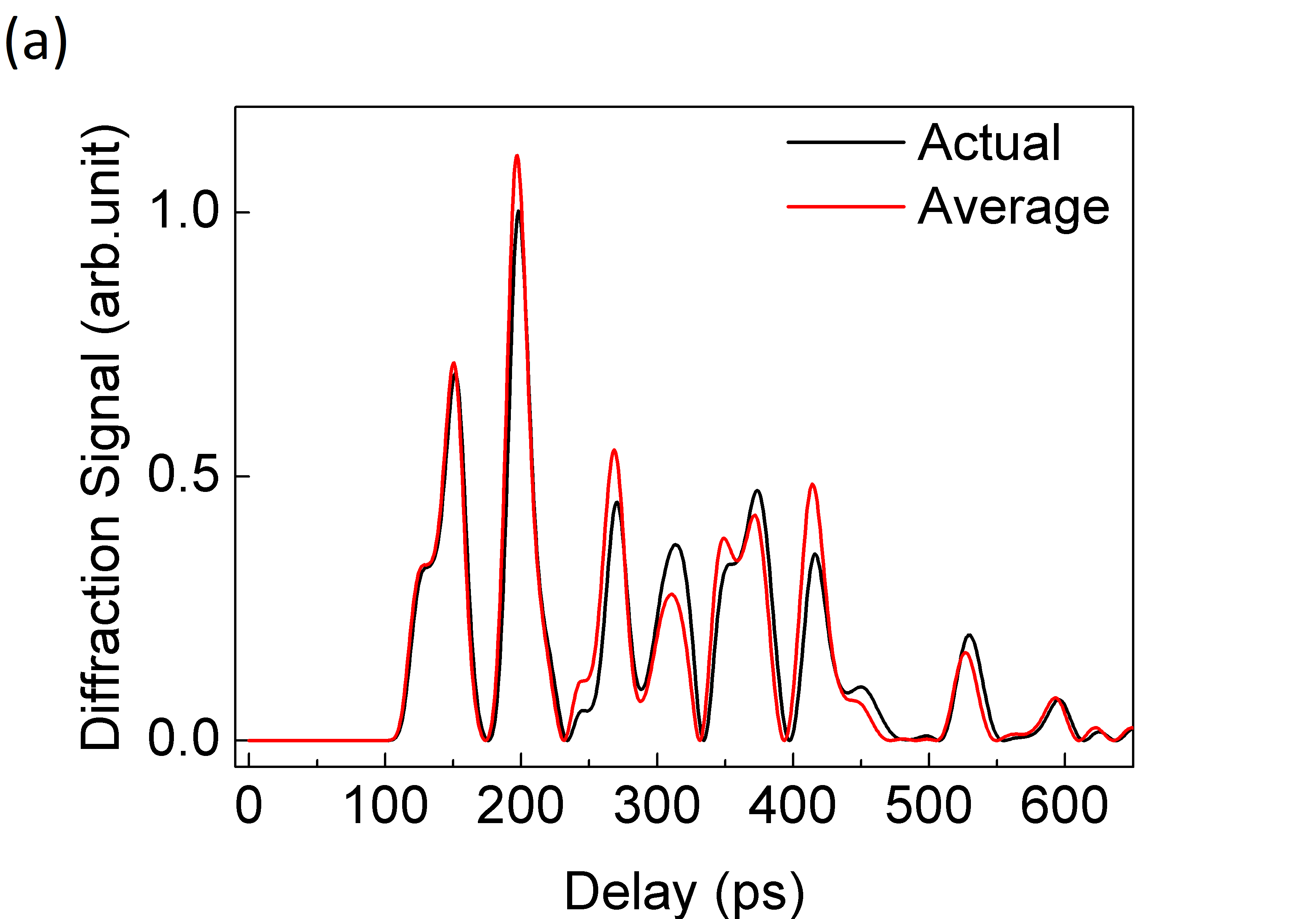}

\includegraphics[width=0.45\textwidth]{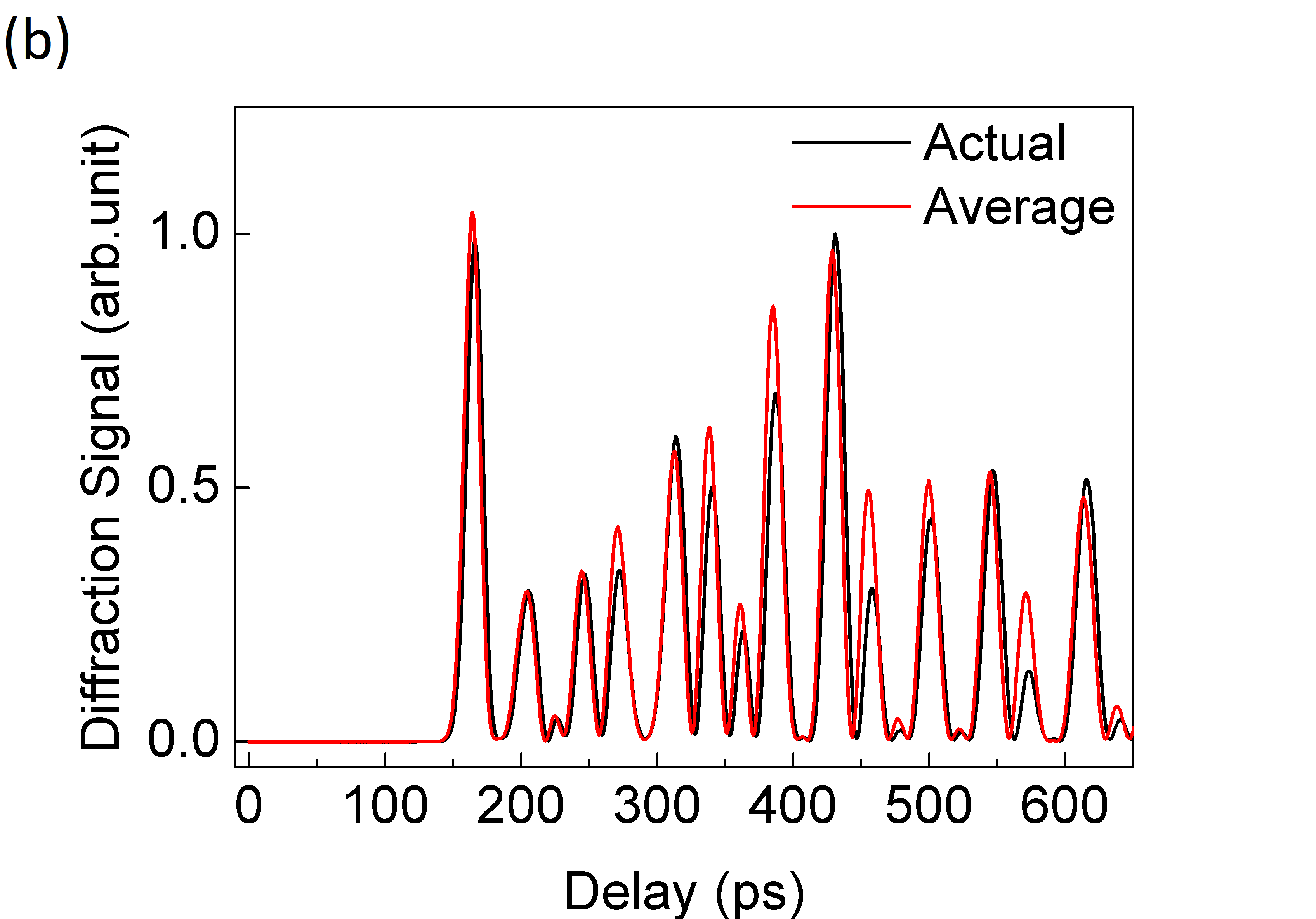}

\caption{(a) Calculated diffraction signal vs. pump-probe delay for the ``Au-multilayer'' sample. The black curve is a calculation that takes all  pairs of alternating layers of 18 nm $\mathrm{SiO_2}$ and 18 nm $\mathrm{Si_3N_4}$ into account. The red curve is a calculation in which the all pairs of alternating layers have been replaced with a single layer having the same total thickness but with average acoustic properties.  (b) Similar to (a) but now for the ``Ni-multilayer'' sample. }

\label{avg_plots}
\end{figure}

To understand how the $\mathrm{SiO_2}$ and $\mathrm{Si_3N_4}$ bilayers affect the strength and shape of the diffraction signal, we performed numerical simulations where we replaced the dielectric bilayers with a single equivalent acoustic medium \cite{ses1}. We replaced the dielectric layers with a single medium that has the same thickness as the $\mathrm{SiO_2}$/$\mathrm{Si_3N_4}$ stack and has a density which is the average density of $\mathrm{SiO_2}$ and $\mathrm{Si_3N_4}$. The acoustic velocity of the equivalent medium is calculated such that the time for the acoustic wave to propagate through the single equivalent medium is the same as the time required to propagate through all the dielectric layers. Hence the velocity of the equivalent time-average medium $(V_{ta})$ is given by 
\begin{equation}
\frac{1}{V_{ta}}= \frac{1}{d_1+d_2} \Big( \frac{d_1}{v_1}+\frac{d_2}{v_2} \Big), 
\end{equation}{}
where $v_1$ and $v_2$ are the acoustic velocities in  $\mathrm{SiO_2}$ and $\mathrm{Si_3N_4}$, respectively and $d_1$ and $d_2$ are the thickness of $\mathrm{SiO_2}$ and $\mathrm{Si_3N_4}$, respectively. In Fig \ref{avg_plots}(a) we show the calculated time-dependent diffraction signal for the  ``Au-multilayer'' sample (black curve), together with the diffraction signal calculated when the 10 dielectric layers are replaced with the single equivalent time-average medium of the same thickness (red curve). Similarly, in Fig \ref{avg_plots}(b) we show the calculated time-dependent diffraction signal for the  ``Ni-multilayer'' sample (black curve) and the calculated diffraction signal when the 20 dielectric layers are replaced with the single equivalent medium of the same thickness (red curve). For both samples, although there are some differences between the calculated curves, the signal shapes are remarkably similar, as are the amplitudes of the signals.  Minor differences are mostly seen at long time delays. The \emph{position} of the first acoustic diffraction peak is not expected to change when the bilayer is replaced with an equivalent acoustic medium because the time it takes for one round trip inside the whole layer remains the same.  However, surprisingly, the partial reflections and transmission at the bilayer interfaces don't change the shape of the time-dependent diffraction signal significantly. This is partially explained by the length of the acoustic wave, which is much greater than the thickness of the individual dielectric layers. The acoustic wave only ``sees'' an equivalent medium rather than the individual layers.  However, the modest acoustic impedance mismatch between the  $\mathrm{SiO_2}$ and $\mathrm{Si_3N_4}$ also plays a role here (see Table 2.), and both effects must be taken into account to understand these results. We note that similar effects have been predicted for the propagation of  low-frequency sound waves in seismology \cite{ses1}. This suggests that laser-induced ultrasonics can be used to detect buried gratings when the number of dielectric layers is further increased to  values often encountered in the semiconductor manufacturing industry.

\subsection{Optical excitation of Au and Ni layers}

\begin{figure}[!]
\centering

\includegraphics[width=0.4\textwidth]{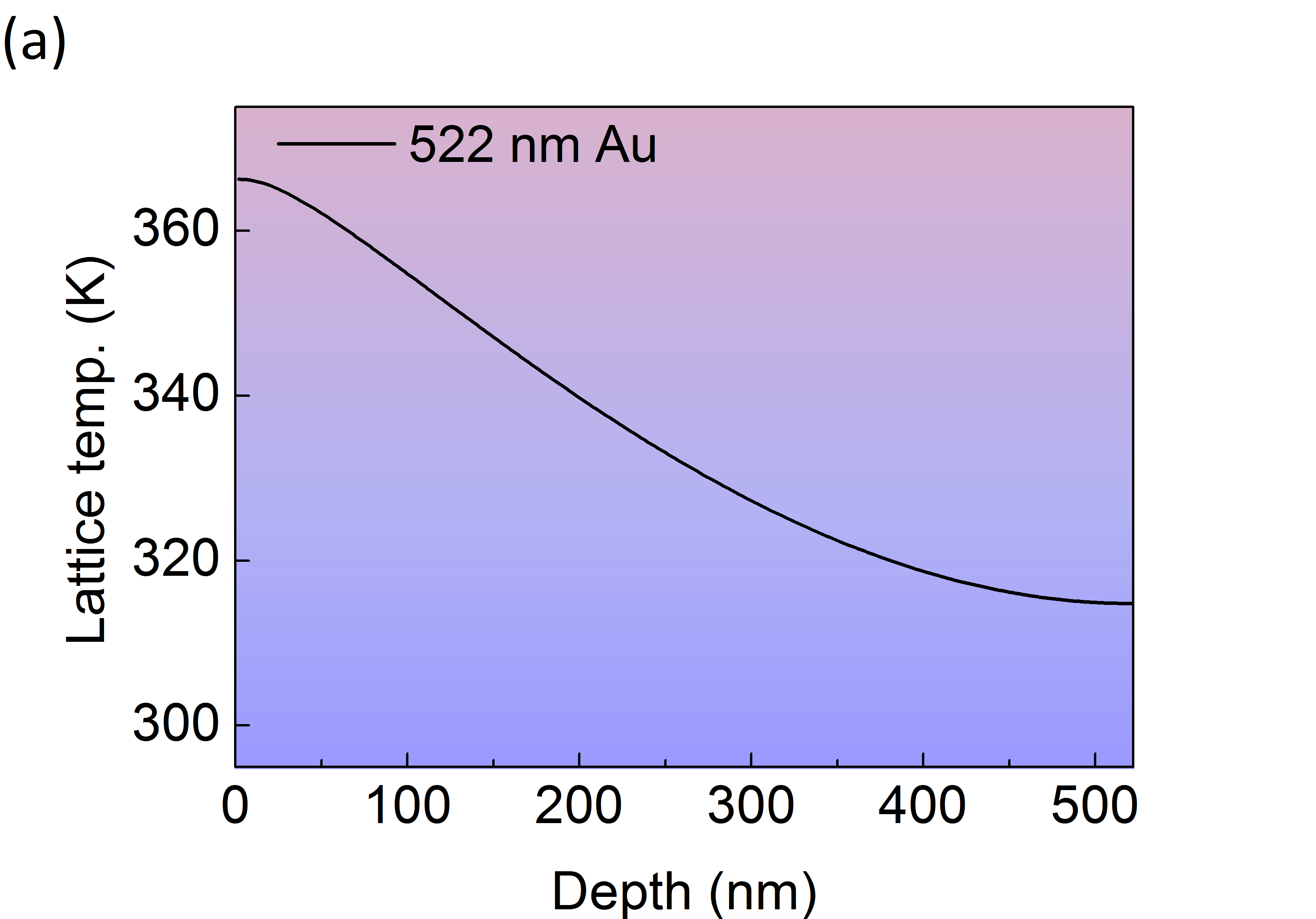}

\includegraphics[width=0.4\textwidth]{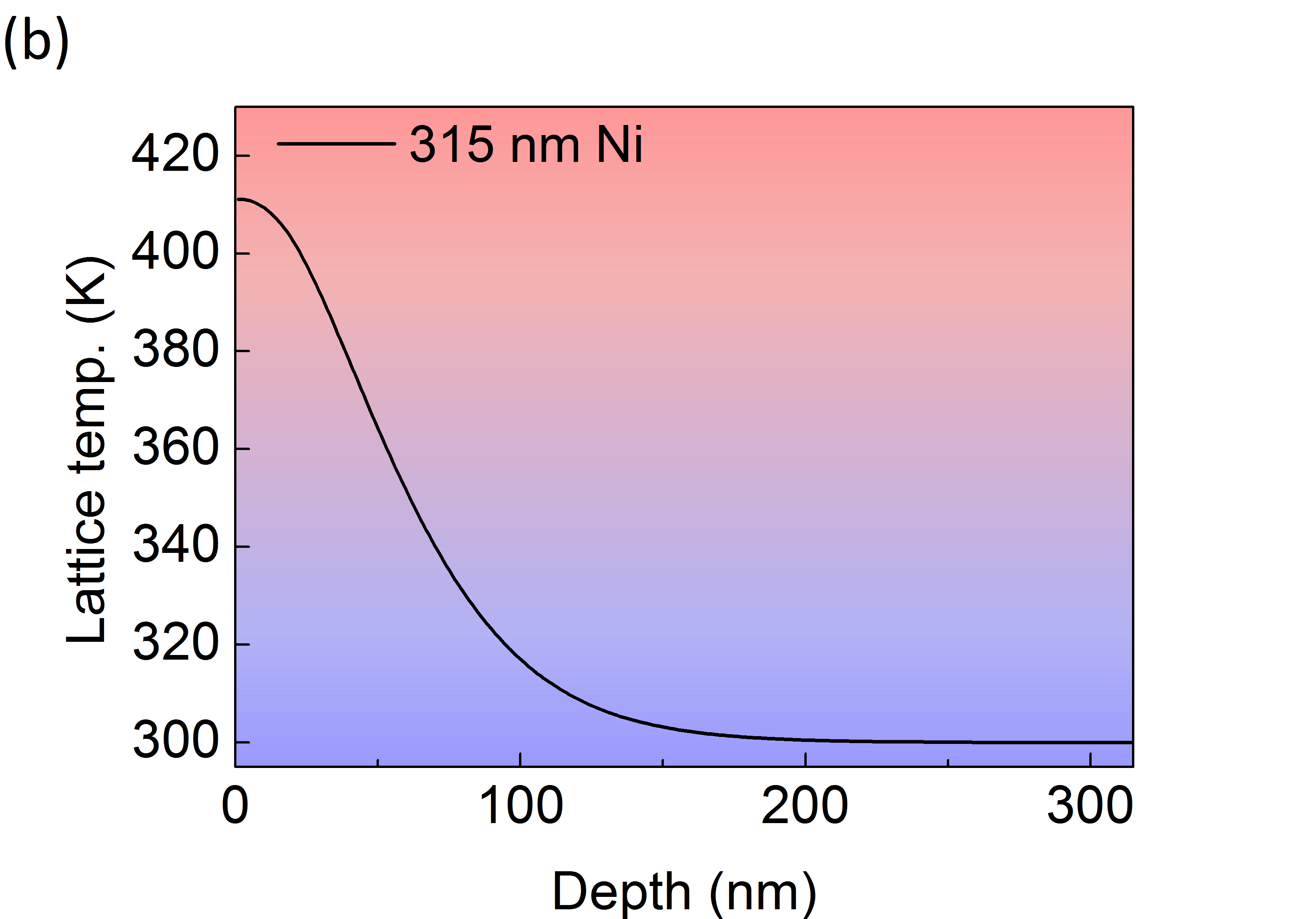}

\includegraphics[width=0.4\textwidth]{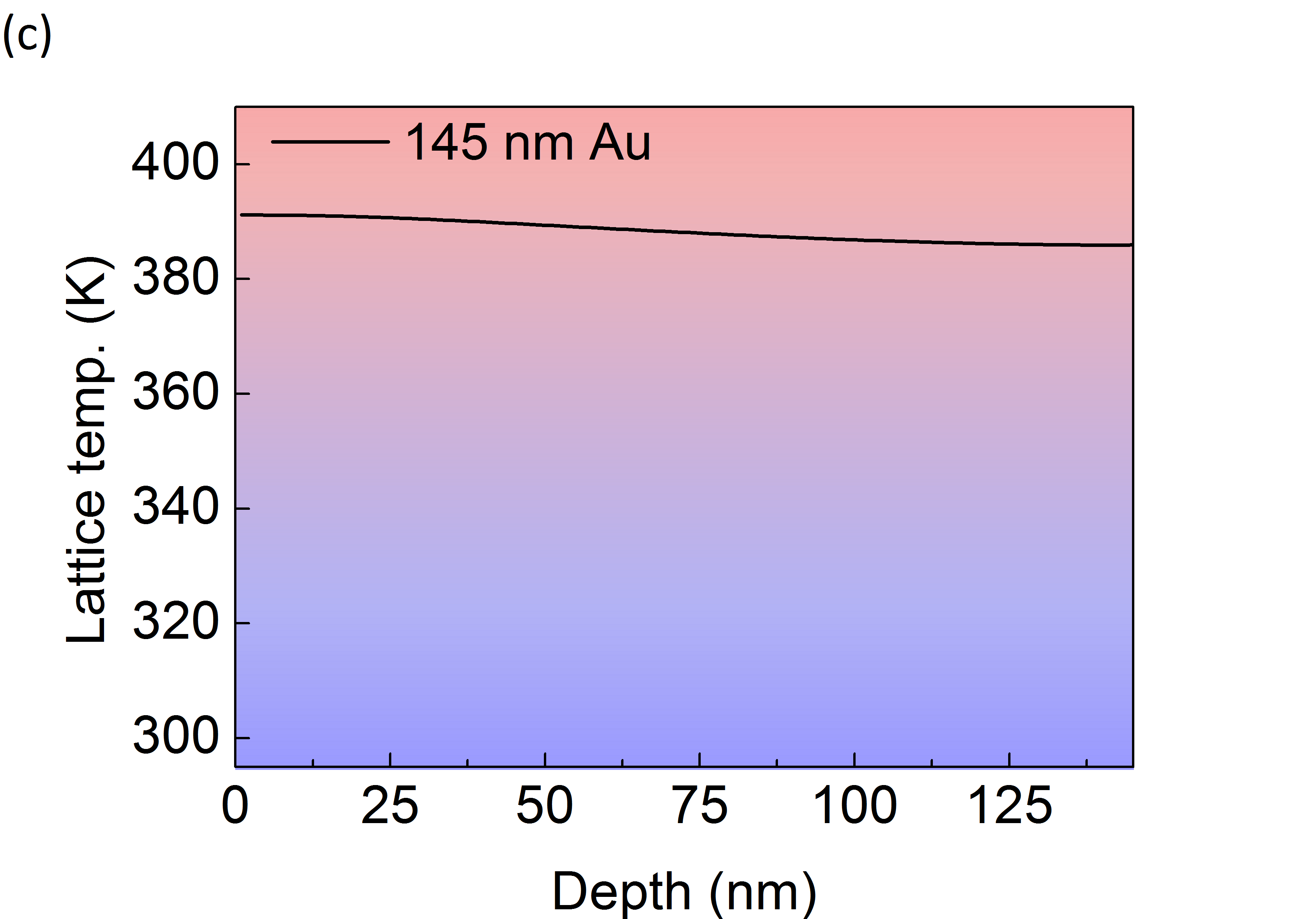}

\includegraphics[width=0.4\textwidth]{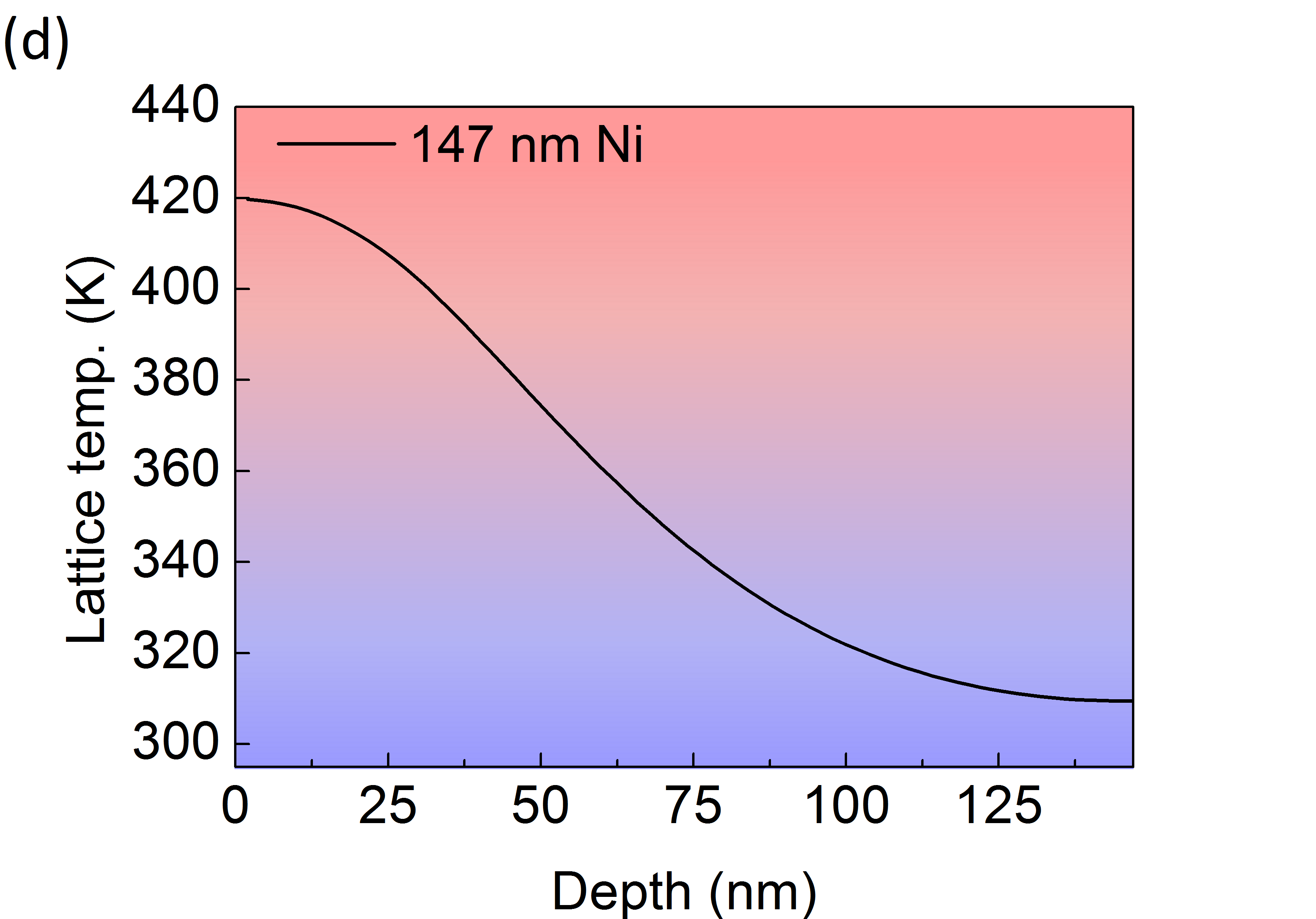}
\caption{ Lattice temperature inside the metal as a function of depth in the metal 15 ps after optical excitation with a 400 nm pump pulse for, (a) a 522 nm thick Au layer, (b) a 315 nm thick Ni layer, (c) a 145 nm Au layer, and (d) a 147 nm Ni layer. }
\label{temp_plot}
\end{figure}

To gain some insight into the longitudinal spatial extent of the acoustic wave generated after optical excitation, we have used the TTM to calculate the lattice temperature as a function of depth inside the metal layer, 15 ps after optical excitation with the pump pulse. At this time the TTM shows that the hot electron gas has significantly cooled and is in local thermal equilibrium with the lattice. In figure  Fig \ref{temp_plot}(a), we show the resulting lattice temperature as a function of depth for the 522 nm Au sample. The figure shows the large penetration depth of energy into the metal, which can be explained by the relatively weak electron-phonon coupling strength of Au.  It is the rapid heating of the lattice that launches the acoustic wave, which, in this case, has a relatively long wavelength. Note that a significant lattice temperature increase is observed even at a depth of 522 nm. 

In Fig \ref{temp_plot}(b), we show the calculated lattice temperature as a function of depth inside the 315 nm thick Ni layer. In contrast to that of Au, the lattice temperature is limited to a smaller depth of about 75 nm, resulting in a higher local temperature and an acoustic wave with a shorter wavelength. The short wavelength of the acoustic wave  manifests itself in our measurements as sharp rising and falling signals in the time-dependent diffraction signal. In Fig \ref{temp_plot}(c), we show the calculated lattice temperature as a function of depth inside the 145 nm thick Au layer present in between the glass and 5 pairs of $\mathrm{SiO_2}$/$\mathrm{Si_3N_4}$. In this case,  the Au layer is thin enough that after 15 ps, the lattice is heated almost homogeneously. There is practically no spatial gradient in the lattice temperature. In Fig \ref{temp_plot}(d), we plot the calculated lattice temperature as a function of depth for the 147 nm thick Ni layer, on which the 10 pairs of $\mathrm{SiO_2}$ and nitride layers have been fabricated.   In contrast to the calculations for 145 nm  Au, here the lattice temperature distribution still shows a significant temperature gradient. The acoustic wave launched by the heated lattice in this case is, therefore, shorter than that in the Au layer.

\subsection{Displacement amplitude at the glass-metal interface}

\begin{figure}[!]
\centering

\includegraphics[width=0.4\textwidth]{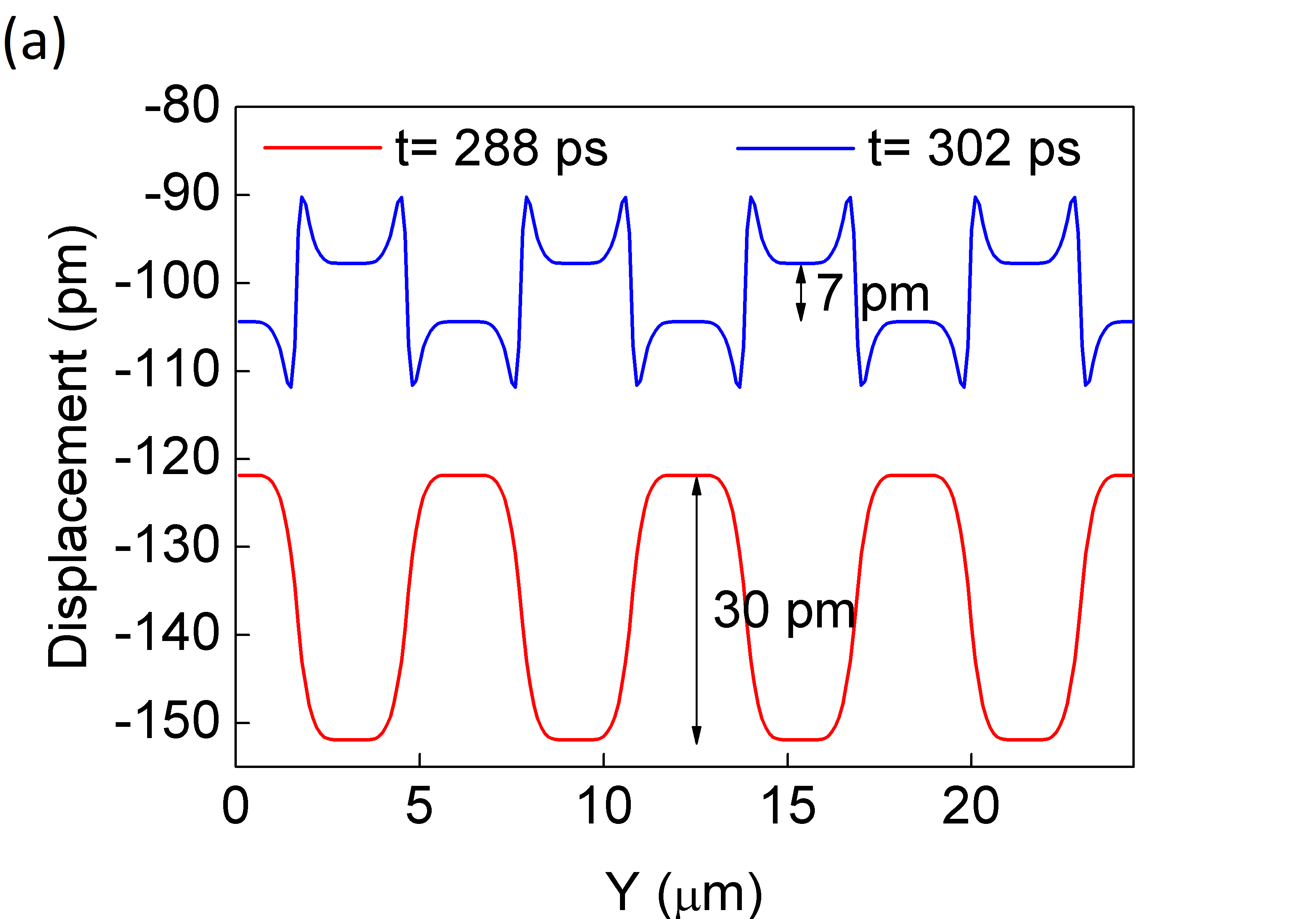}

\includegraphics[width=0.4\textwidth]{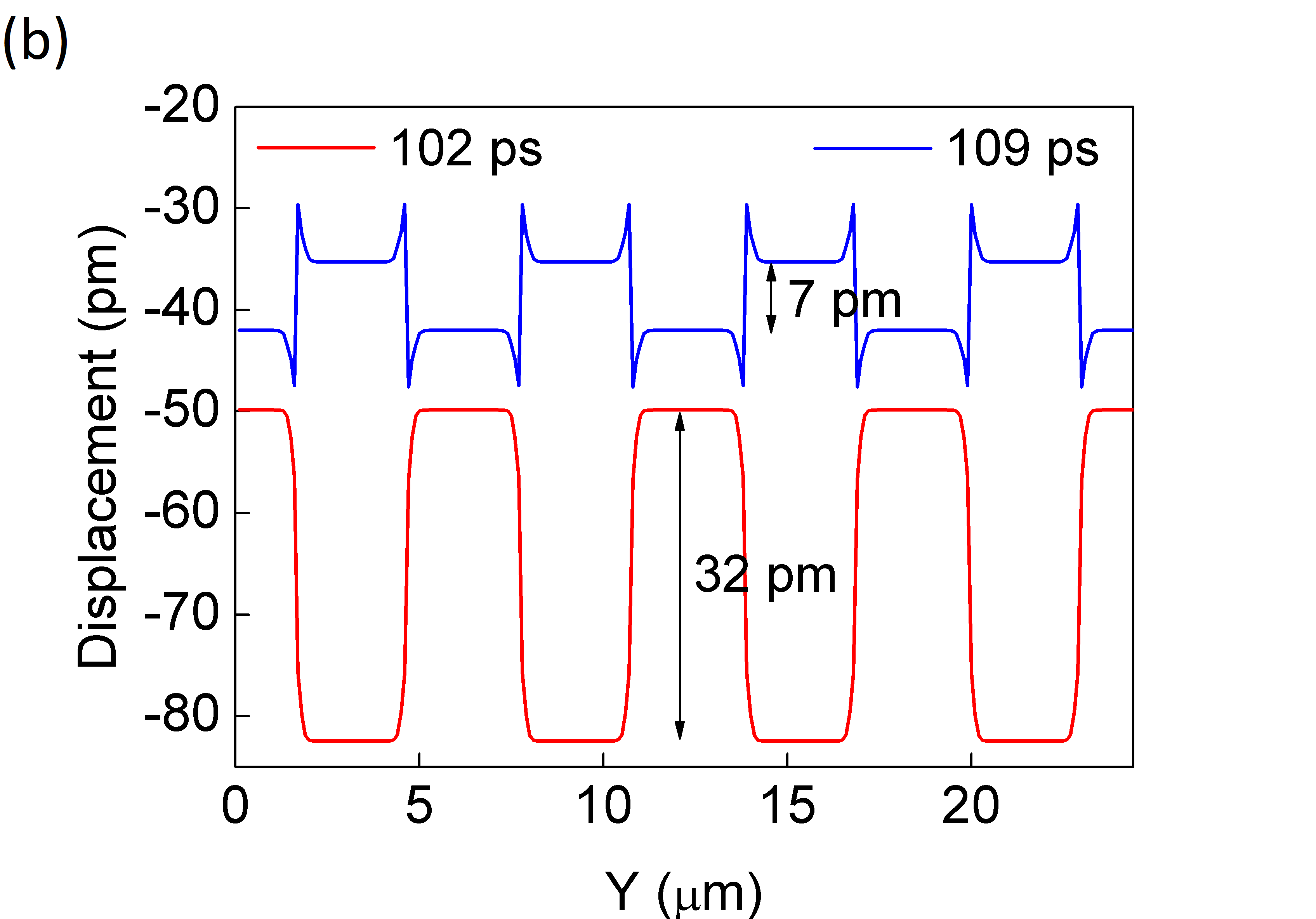}

\includegraphics[width=0.4\textwidth]{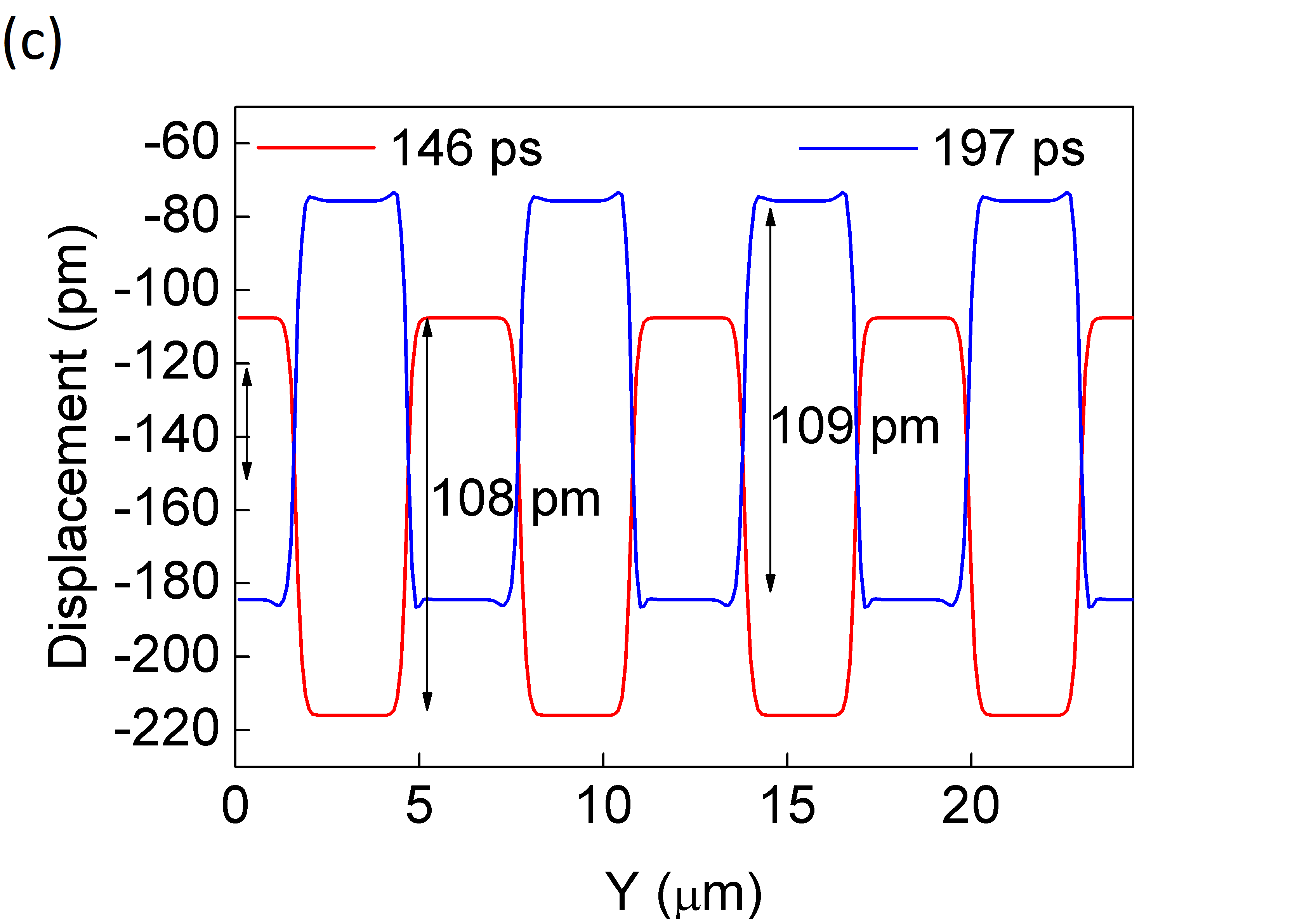}

\includegraphics[width=0.4\textwidth]{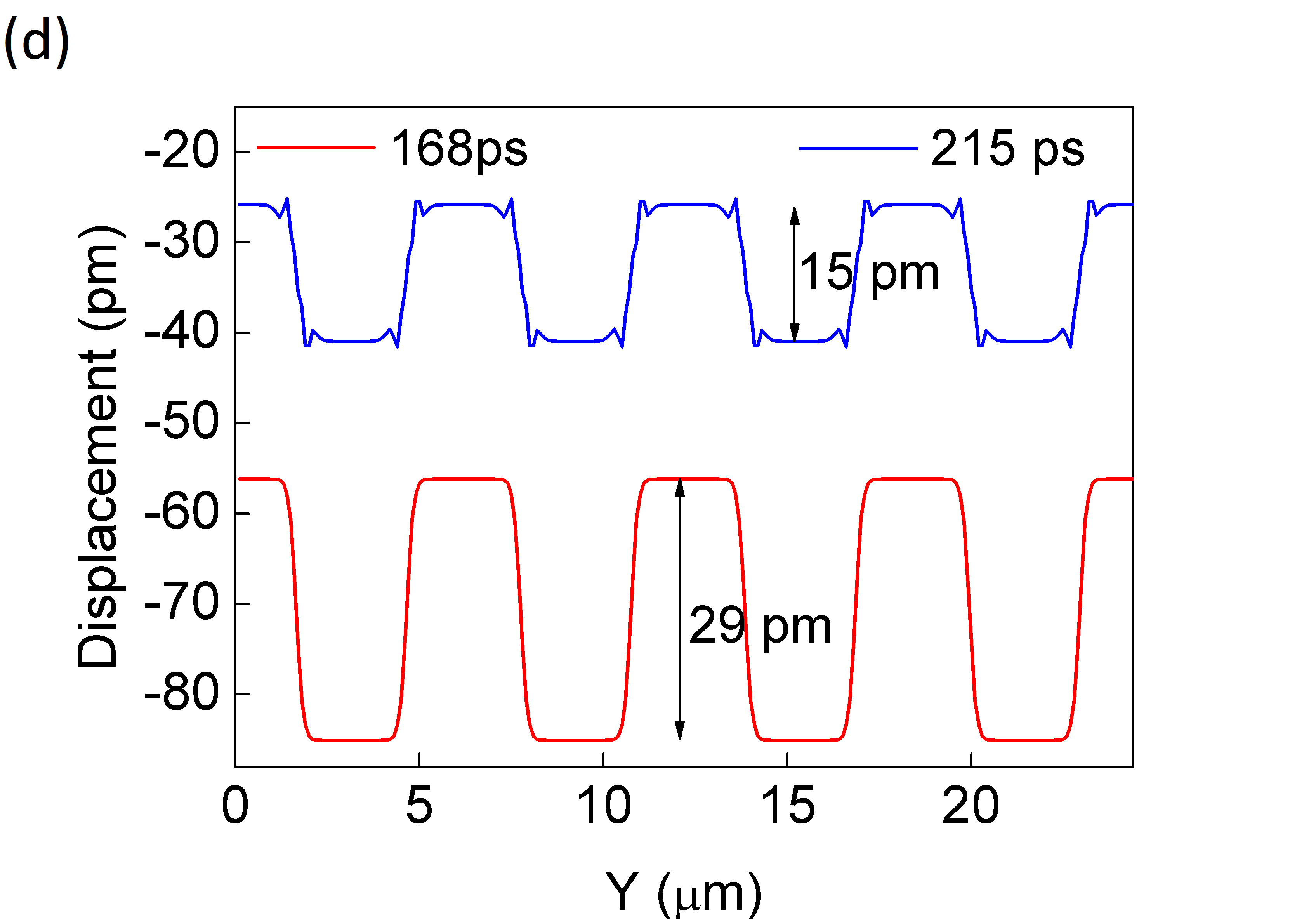}

\caption{ (a) Lattice displacement calculated at the glass-Au interface for the 522 nm Au sample with the 10 nm buried grating at time delays of 288 ps and 302 ps after optical excitation.  (b) Lattice displacement calculated at the glass-Ni interface for the 315 nm Ni sample with 10 nm buried grating at time delays of 102 ps and 109 ps after optical excitation. (c) Lattice displacement calculated at the glass-Au interface for the ``Au-multilayer'' sample at time delays of 146 ps and 197 ps after optical excitation. (d) Lattice displacement calculated at the glass-Ni interface for the ``Ni-multilayer'' sample at time delays of 168 ps and 215 ps after optical excitation.}
\label{dis_plot}
\end{figure}

To get an estimate of the typical surface displacement amplitudes near the glass-metal interface, we plot in Fig \ref{dis_plot} the lattice displacement as a function of the position along the direction perpendicular to the grating lines for two different times after optical excitation, for the four different samples discussed in this paper. Here, four unit cells used in the simulations are shown for clarity, where one unit cell has a width of 6 $\mathrm{\mu m}$. In the convention used here, the unperturbed glass-metal interface has zero amplitude displacement, and a negative displacement implies a movement of the interface  in the direction of the glass substrate.  In Fig \ref{dis_plot}(a), we plot the  displacement of the glass-Au interface for the 522 nm Au sample with the 10 nm buried grating, for pump-probe delays of 288 ps (red curve) and 302 ps (blue curve).  At 288 ps, the acoustic wave has completed one round trip inside the 522 nm Au, and the maximum diffraction efficiency due to the displacement of the Au lattice is calculated. The grating-shaped displacement at the glass-Au interface has the same phase as  the buried grating and has a peak-to-valley amplitude of 30 pm. The lattice displacement  shows an offset of over 100 pm, which is partially caused by the acoustic wave and partially by the expansion of the lattice due to heating. However, only the grating-shaped displacement profile contributes to the diffraction efficiency. At 302 ps, we observe that the phase of the grating-shaped lattice displacement has changed by $\pi$ and the peak-to-valley amplitude has decreased to 7 pm. The $\pi$ phase change  can be explained by the presence of two acoustic-wave gratings: one reflected off the valleys of the buried grating, the other off the peaks of the buried grating. The interference between the acoustic wave gratings, which have slightly different arrival times, leads to the occasional inversion of the interface displacement grating, which is equivalent to a spatial $\mathrm{\pi}$ phase change.

A similar effect can be seen in the lattice displacement plot for the 315 nm Ni sample shown in Fig \ref{dis_plot}(b). The grating-shaped displacement at 102 ps (red curve) has the same phase as  the buried grating and the $\pi$ phase shift of the displacement grating occur at 109 ps (blue curve). Another interesting observation from these calculations is the difference in the shape of the lattice displacement grating for the Au and Ni samples shown by the red curves in Fig \ref{dis_plot}(a) and Fig \ref{dis_plot}(b), respectively. For the Ni sample, the grating-shaped displacement profile is `sharper' than for the Au sample. This difference is  due to acoustic diffraction from the buried grating. Since the acoustic waves generated inside the Au layer have a much larger wavelength, acoustic wave diffraction is more prominent than inside the Ni layer.  This  leads to a more smeared out, smoother grating at the interface.

In Fig \ref{dis_plot}(c), we show the displacement at the glass-Au interface for the ``Au-multilayer'' sample at pump-probe delays of 146 ps (red curve) and 197 ps (blue curve). The grating-shaped displacement at 146 ps has the same phase as  the buried grating and has a peak-to-valley amplitude of 108 pm. The displacement amplitude is larger than in the case of the 522 nm Au sample because the lattice heating in case of 145 nm Au is confined to a smaller volume which in turn gives rise to a stronger acoustic wave. This is also evident from the heating profile curves in Fig \ref{temp_plot}(a) and  Fig \ref{temp_plot}(c).  The grating-shaped displacement at 197 ps has a slightly higher peak-to-valley amplitude of 109 pm and, hence, a higher diffraction efficiency. 

In Fig \ref{dis_plot}(d) we show the displacement at the glass-Ni interface for the ``Ni-multilayer'' at a pump-probe delay of 168 ps, when the acoustic wave has completed one round trip (red curve). The grating-shaped displacement profile has a peak-to-valley amplitude of 29 pm and has the same phase as that of the buried grating. The grating-shaped acoustic wave, now at the glass-Ni interface, undergoes another reflection inside  the 147 nm Ni layer. The acoustic wave returns to the glass-Ni interface and results in optical diffraction  215 ps after optical excitation.  Hence, the grating-shaped displacement at 215 ps (blue curve in Fig \ref{dis_plot}(d) ), has the same phase as the one at 168 ps but a reduced peak-to-valley amplitude of 15 pm. This  displacement grating has a lower amplitude  because  the acoustic wave is also transmitted into the stack of dielectric layers upon reflection at the Ni-$\mathrm{SiO_2/Si_3N_4}$ interface,  and also due to damping and dispersion of the acoustic wave during the propagation through the 147 nm Ni layer. 

\subsection{Material properties}

\begin{table}[h!]
\begin{tabular}{ cccc } 
 \hline
   \hline
  & Au & Ni  \\ 
  \hline

 Optical penetration depth at 400 nm (nm)  & 16 & 12 \\ 

 Optical penetration depth at 800 nm (nm) & 13 & 13 \\ 

  Electron-phonon coupling constant ($\mathrm{10^{16} Wm^{-3}K^{-1}}$) & 3.2 & 36 \\ 
 \hline
   \hline
 \end{tabular}
 \caption{Properties of the Au and Ni used in our TTM calculations\cite{aug1,aug2,aupd,nipd,ed2}.}
\label{optical_ppt}
\end{table}

\begin{table}[h!]
\centering
\begin{tabular}{ ccc } 
  \hline
    \hline
& Sound velocity    &  Acoustic Impedance    \\
&  (m/s)   &   ($\mathrm{10^6 Ns/m^3}$)    \\
  \hline
  Au & 3,200 & 63.8 \\

  Ni & 5,800 & 51.5 \\

  Glass substrate & 5,700 & 12.54 \\

  $\mathrm{SiO_2}$ & 5,100 & 14.8 \\
  
   $\mathrm{Si_3N_4}$ & 5,600 & 17.9\\
 \hline
  \hline
 \end{tabular}
 \caption{Acoustic properties of different materials used in our calculations\cite{ac28, ac5}.}
\label{acoustic_ppt}
\end{table}

\section{Conclusion}

We have shown that laser-induced ultrasonics can be used to detect the presence of gratings buried under optically opaque metal and dielectric layers.We observe optical diffraction from the acoustic wave reflected from a grating buried under thick Au and Ni layers. The diffraction is due to the grating-shaped displacement of atoms at the glass-metal interface and Brillouin scattering in the glass substrate.  We attribute the difference in the shape of the time-dependent diffraction signal for Au and Ni to the difference in electron-phonon coupling strength of these two metals. Our measurements on complex multilayer samples show that the acoustic wave can be detected even after it has propagated through multiple  $\mathrm{SiO_2}$ and $\mathrm{Si_3N_4}$ layers. The numerical calculations are in agreement with the measurements and also show that the acoustic wave ``sees' the  $\mathrm{SiO_2}$/$\mathrm{Si_3N_4}$ stack as an equivalent time-averaged acoustic medium. The results of our experiments and simulations strongly suggest that this technique can be used for sub-surface metrology applications, especially in the semiconductor device manufacturing industry.

\begin{acknowledgments}
This work is carried out at the Advanced Research Center for Nanolithography (ARCNL), a public-private partnership of the University of Amsterdam (UvA), the Vrije Universiteit Amsterdam (VU), the Netherlands Organisation for Scientific Research (NWO), and the semiconductor equipment manufacturer ASML.

\end{acknowledgments}


\bibliography{main}

\end{document}